\documentclass[aps,prd,preprint,onecolumn,citeautoscript,superscriptaddress,nofootinbib,eqsecnum]{revtex4-1}  
\usepackage{subfig}
\usepackage{graphicx}
\usepackage{amsmath,amssymb,amsthm,amsfonts}
\usepackage{multirow,array,bm,bbm,esint}
\usepackage{epsf}
\usepackage{float}
\usepackage{slashed}
\usepackage[papersize={8.5in,11in}]{geometry}
\usepackage[colorlinks=true]{hyperref}  
\usepackage{comment}

\usepackage{color}

\hypersetup{
    bookmarks=true,         
    unicode=false,          
    pdftoolbar=true,        
    pdfmenubar=true,        
    pdffitwindow=false,     
    pdfstartview={FitH},    
    pdftitle={},    
    pdfauthor={},     
    pdfsubject={},   
    pdfcreator={Creator},   
    pdfproducer={Producer}, 
    pdfkeywords={} {} {}, 
    pdfnewwindow=true,      
    colorlinks=true,       
    linkcolor=purple!100!red!100, 
    citecolor=blue,        
    filecolor=magenta,      
    urlcolor=blue           
} 

\newcommand{\cD}{{\cal D}}
\newcommand{\psib}{{\overline{\psi}}}
\newcommand{\Qb}{{\overline{Q}}}
\newcommand{\cO}{{\cal O}}

\def\hf{\frac{1}{2}}
\def\qtr{\frac{1}{4}}

\def\nn{\nonumber}
\def\bec{\begin{center}}
\def\eec{\end{center}}
\def\beq{\begin{equation}}
\def\eeq{\end{equation}}
\def\bseq{\begin{subequations}}
\def\eseq{\end{subequations}}
\def\bea{\begin{eqnarray}}
\def\eea{\end{eqnarray}}
\usepackage{todonotes}

\begin{document}

\title{Probing Non-perturbative Supersymmetry Breaking through Lattice Path Integrals}

\author{Navdeep Singh Dhindsa}
\email{navdeep.s.dhindsa@gmail.com}

\author{Anosh Joseph}
\email{anoshjoseph@iisermohali.ac.in}

\affiliation{Department of Physical Sciences, Indian Institute of Science Education and Research (IISER) Mohali, Knowledge City, Sector 81, SAS Nagar, Punjab 140306, India}

\date{\today
\\
\vspace{0.8in}}

\begin{abstract}
\vspace{0.2in}

We investigate non-perturbative supersymmetry breaking in various models of quantum mechanics, including an interesting class of $PT$-invariant models, using lattice path integrals. These theories are discretized on a temporal Euclidean lattice with anti-periodic boundary conditions. Hybrid Monte Carlo algorithm is used to update the field configurations to their equilibrium values. We used the Ward identities and expectation values of superpotentials as tools for probing supersymmetry breaking. 

\end{abstract}


\maketitle

\tableofcontents

\section{Introduction}

We can use supersymmetric quantum mechanics as a testbed to illustrate several properties of systems containing bosons and fermions. Since Witten's seminal work \cite{Witten:1981nf}, the idea of non-perturbative supersymmetry (SUSY) breaking has been investigated extensively in the literature. These investigations range from studying the properties of supersymmetric quantum mechanics to supersymmetric gauge theories in various spacetime dimensions \cite{Shifman:1995ua, Izawa:1996pk, Shadmi:1999jy, Kanamori:2007yx, Dine:2010cv, Wozar:2011gu, Kadoh:2015zza, Catterall:2017xox, Kadoh:2018ivg, Kadoh:2019bir}. In this work, we investigate non-perturbative SUSY breaking in various quantum mechanics models by regularizing them on a Euclidean lattice. Supersymmetric quantum mechanics models have been the subject of thorough investigations in the context of various physical systems over the past few decades (see Ref. \cite{FernandezC:2018cdo} for a review). For example, the model with a quartic superpotential (supersymmetric anharmonic oscillator) has been simulated on the lattice, by several groups, over the past ten years or so, with great success (see Refs. \cite{Catterall:2000rv, Giedt:2004vb, Kanamori:2007ye, Bergner:2007pu, Schierenberg:2012pb, Baumgartner:2014nka, Baumgartner:2015qba}). In this work we explore supersymmetric quantum mechanics with various types of superpotentials, including the interesting class of $PT$-invariant potentials. After verifying the existing simulation results in the literature on supersymmetric anharmonic oscillator, with the help of a lattice regularized action, and an efficient simulation algorithm, we use the same set up to probe SUSY breaking in models with three different superpotentials. They include a degree-five potential, a shape invariant potential of Scarf I type, and a certain type of $PT$-invariant potential. Although the Euclidean action of the $PT$-symmetric quantum mechanics we consider here can be complex, for an arbitrary value of the parameter $\delta$ appearing in the potential, we were able to simulate a subset of it; for $\delta = 0, 2, 4$, where the action becomes real. Our simulations indicate that SUSY is preserved in these models, in agreement with the recent simulations based on complex Langevin method \cite{Joseph:2020gdh}. The models discussed in this work were simulated using lattice regularized versions of the Euclidean path integrals, with the Hybrid Monte Carlo (HMC) algorithm to update the field configurations in simulation time. We note that for general $PT$-symmetric models with complex actions we can use complex Langevin method to explore non-perturbative SUSY breaking. (See \cite{Joseph:2019sof, Joseph:2020gdh} for some recent results.)

The plan of this paper is as follows. In Sec. \ref{sec:Supersymmetric_Quantum_Mechanics} we briefly discuss supersymmetric quantum mechanics in the continuum and also on a lattice. In Sec. \ref{sec:Supersymmetric Anharmonic Oscillator} we cross-check our simulation results with the ones existing in the literature for supersymmetric anharmonic oscillator. In Sec. \ref{sec:Model_with_Odd-Degree_Superpotential} we study the model with degree-five superpotential. Our simulation results indicate that SUSY is dynamically broken in this model, which is in agreement with the analytical result in Ref. \cite{Witten:1981nf}. In Sec. \ref{sec:Scarf_I_Superpotential} we study the model with a particular type of shape invariant potential known as the Scarf I potential. Our simulations indicate that this model also has an intact SUSY, and thus confirm the numerical results given in Ref. \cite{Kadoh:2018ele}. In Sec. \ref{sec:SUSYQM_with_PT-Symmetry} we study the possibility of SUSY breaking in certain models with $PT$-invariant superpotentials. Our simulation results point to intact SUSY in these models as well. The conclusions on $PT$-invariant models also agree with the recent simulation results obtained using complex Langevin method \cite{Joseph:2020gdh}. We provide conclusions and future directions in Sec. \ref{sec:Conclusions}. 

\section{Supersymmetric Quantum Mechanics}
\label{sec:Supersymmetric_Quantum_Mechanics}

\subsection{Continuum Theory}
\label{sec:Continuum_Theory}

The Euclidean action of supersymmetric quantum mechanics has the off-shell form
\beq
\label{eq:model-action}
S = \int_0^\beta d\tau \left( - \hf \phi \partial_\tau^2 \phi + \psib \partial_\tau \psi - \hf B^2 + \psib W''(\phi) \psi - B W'(\phi) \right).
\eeq
Here, the integral is over a compactified time circle of circumference $\beta$ in Euclidean time. The fields $\phi$ and $B$ are bosonic while $\psib$ and $\psi$ are fermionic. They depend only on the Euclidean time variable $\tau$. The derivative with respect to $\tau$ is denoted as $\partial_\tau$. The superpotential $W(\phi)$ completely determines the interactions in the theory. The primes denote the derivatives of superpotential with respect to $\phi$. 

The above action is invariant under two supercharges $Q$ and $\Qb$. They obey the following algebra
\beq
\label{eq:algebra}
\{ Q, Q \} = 0, ~~ \{\Qb, \Qb \} = 0, ~~ \{ Q, \Qb \} = 2 \partial_\tau.
\eeq

The auxiliary field $B$ appearing in the action can be integrated out using its equation of motion $B = - W'(\phi)$. Then the on-shell action takes the form
\beq
S = \int d\tau \left( - \hf \phi \partial_\tau^2 \phi + \psib \partial_\tau \psi + \psib W''(\phi) \psi + \hf \left[ W'(\phi) \right]^2 \right). 
\eeq

The supercharges $Q$ and $\Qb$ act on the fields in the following way
\beq
Q \phi = \psib, ~~~ Q \psi = - \partial_\tau \phi + W', ~~ Q \psib = 0,
\eeq
and
\beq
\Qb \phi = - \psi, ~~ \Qb \psi = 0, ~~ \Qb \hspace{0.05cm} \psib = \partial_\tau \phi + W'.
\eeq

Since SUSY is not broken explicitly in our models, we are interested in detecting the presence or absence of SUSY breaking that can arise from non-perturbative (dynamical) effects. In order to probe this, we numerically simulate these models on a lattice. Since lattice formulation of a quantum theory is inherently non-perturbative we can detect SUSY breaking by simulating the model using a suitable Monte Carlo algorithm.

The partition function of the model is 
\beq
\label{eqn:cont-pf}
Z = \int \cD \phi \cD \psi  \cD \psib ~e^{-S[\phi, \psi, \psib]}
\eeq
with periodic temporal boundary conditions for all the fields. 

Consider the case when SUSY is preserved in the system. Then the Hamiltonian $H$, corresponding to the Lagrangian in Eq. \eqref{eq:model-action}, with energy levels $E_n$, $n = 0, 1, 2, \dots$, will have the ground state energy $E_0 = 0$. The bosonic and fermionic excited states
\beq
| b_{n+1} \rangle = \frac{1}{\sqrt{2 E_{n+1}}} \bar{Q} |f_n \rangle, ~~~|f_n \rangle = \frac{1}{\sqrt{2 E_{n+1}}} Q |b_{n+1} \rangle
\eeq
form a SUSY multiplet satisfying the algebra Eq. \eqref{eq:algebra}. In the above, $|b_0 \rangle$ is the ground state of the system. Assuming that the states $|b_n \rangle$ and $|f_n \rangle$ have the fermion number charges $F = 0$ and $F = 1$, respectively, when periodic temporal boundary conditions are imposed for both the bosonic and fermionic fields, the partition function Eq. \eqref{eqn:cont-pf} is equivalent to the Witten index ${\cal W}$ \cite{Witten:1982df}. We can see that 
\beq
Z  \equiv {\cal W} = {\rm Tr} \left[ (-1)^F e^{-\beta H} \right] 
=  \langle b_0 | b_0 \rangle + \sum_{n = 0}^\infty \left[ \left(\langle b_{n+1} | b_{n+1} \rangle - \langle f_n | f_n \rangle  \right) e^{-\beta E_{n+1}} \right] 
\eeq
does not vanish due to the existence of a normalizable ground state. This in turn makes the normalized expectation values of observables well-defined. 

In the SUSY broken case, we end up in a not-so-trivial situation. The Hamiltonian $H$ corresponding to the Lagrangian in Eq. \eqref{eq:model-action} has a positive ground state energy ($0 < E_0 < E_1 < E_2 \dots$). The SUSY multiplet is defined as
\beq
| b_n \rangle = \frac{1}{\sqrt{2 E_n}} \bar{Q} | f_n \rangle, ~~~| f_n \rangle = \frac{1}{\sqrt{2 E_n}} Q | b_n \rangle.
\eeq

Differently from the unbroken SUSY case, when SUSY is broken, the supersymmetric partition function
\beq
Z  \equiv {\cal W} = {\rm Tr} \left[ (-1)^F e^{-\beta H} \right] = \sum_{n = 0}^\infty \left[ \left( \langle b_n | b_n \rangle - \langle f_n | f_n \rangle  \right) e^{-\beta E_n} \right]
\eeq
vanishes due to the cancellation between bosonic and fermionic states.  As a consequence, the normalized expectation values of observables will be ill-defined. 

In order to avoid this difficulty, we can consider the system at non-zero temperature, to examine the question of SUSY breaking. Thermal boundary conditions explicitly break SUSY. However, we can take the zero temperature limit, and in this limit if $E_0 > 0$, we infer that non-perturbative SUSY breaking occurs in the model.

We make use of three different observables to probe SUSY breaking. These observables are the Ward identities, expectation value of the action, and the expectation value of the first derivative of the superpotential. In a model with unbroken SUSY, the Ward identities should fluctuate around zero in the middle region of the lattice. The expectation value of the action should behave in a certain way (which we will see later) when SUSY is preserved. The expectation value of the first derivative of the superpotential should be zero in a model with intact SUSY. Thus, these three different observables as a whole will help us in identifying the absence or presence of SUSY breaking in a reliable manner.    

We can derive a Ward identity by rewriting the expectation value of an observable say, $\cO(\phi)$ by considering the infinitesimal transformations $\phi \to \phi' = \phi + \delta \phi$ with $\cD \phi' = \cD \phi$. Under these transformations $\langle \cO \rangle$ becomes
\bea
\langle \cO \rangle &=& \frac{1}{Z} \int \cD \phi' \cO(\phi') e^{-S(\phi')} \nn \\
&=& \frac{1}{Z} \int \cD \phi \cO(\phi) e^{-S(\phi)} [ 1 - \delta S(\phi) ] + \frac{1}{Z} \int \cD \phi \delta \cO (\phi) e^{-S(\phi)} [ 1 - \delta S(\phi) ].
\eea

In the above, we have expanded the exponential up to first order. Upon neglecting the $\delta S \delta \cO $ term we get
\bea
\langle \cO \rangle &=& \langle \cO \rangle - \frac{1}{Z} \int \cD \phi \cO(\phi) \delta S(\phi) e^{-S(\phi)} + \frac{1}{Z} \int \cD \phi \delta \cO(\phi) e^{-S(\phi)}.
\eea

Thus we are left with $\langle \cO \delta S \rangle = \langle \delta \cO \rangle$. Since the action is invariant under the infinitesimal transformation $\delta$, we must have $\langle \delta \cO \rangle = 0$. This results in the Ward identity
\beq
\langle \cO \delta S \rangle = 0.
\eeq
If this relation is respected in our numerical simulations then we have a SUSY preserving theory.

\subsection{Lattice Theory}
\label{sec:Lattice_Theory}

Let us consider path integral quantization of the model, discretized on a one-dimensional Euclidean lattice. The lattice has $N_{\tau}$ number of equally spaced sites with the lattice spacing $a = \beta N_{\tau}^{-1}$, where $\beta$ is inverse temperature. The lattice action has the form
\beq
\label{eq:action}
S = S_B + \sum_{i j} \psib_i M_{ij} \psi_j,
\eeq
where the bosonic part of the action is
\beq
\label{eq:s_B}
S_B = \sum_{ij} \left( - \hf \phi_i D_{ij}^2 \phi_j \right) + \hf \sum_i W'_i W'_i.
\eeq
The indices $i$ and $j$ represent the lattice sites and they run from $0$ to $N_{\tau}-1$. The fermion operator is denoted as $M$ and the elements of this matrix, $M_{ij}$, connect the sites $i$ and $j$. The superpotential at site $i$ is denoted by $W_i$. 

It is well known that for supersymmetric theories on a lattice the presence of fermion doublers breaks SUSY. We could include a Wilson-mass term in the model to remove the doublers \cite{PhysRevD.59.054507} and to get the expected target theory in the continuum. The Wilson-mass matrix $K_{ij}$ has the following form
\beq
K_{ij} = m \delta_{ij} - \hf \left( \delta_{i, j+1} + \delta_{i, j-1} - 2 \delta_{ij} \right).
\eeq

Upon defining $W''_{ij}$ as
\beq
W''_{ij} \equiv K_{ij} + W'_i \delta_{ij},
\eeq
and the symmetric difference operator
\beq
\label{d}
D_{ij} \equiv \hf \left( \delta_{j, i+1} - \delta_{j, i-1} \right),
\eeq
we can express the elements of the fermionic matrix as
\beq
\label{eq:ferm_matrix}
M_{ij} = D_{ij} + W''_{ij}.
\eeq

In order to simulate the fermionic sector of the theory we replace the fermions by pseudo-fermions $\chi$ \cite{Kennedy:2006ax}. Then the action in Eq. \eqref{eq:action} takes the form
\beq
S = S_B + S_F.
\eeq

Expressing the bosonic action $S_B$, given in Eq. \eqref{eq:s_B}, with the help of the symmetric difference operator defined in Eq. \eqref{d} we get
\beq
S_B = \sum_i \left[ -\frac{1}{8} \left( \phi_i \phi_{i+2} + \phi_i \phi_{i-2} - 2 \phi_i^2 \right) + \hf W'_i W'_i \right].
\eeq

The fermionic part of the action takes the following form in terms of the pseudo-fermion fields
\beq
S_F = \sum_{ij} \Big( \chi_i (M^T M)^{-1}_{ij} \chi_j \Big).
\eeq

Let us look at the Ward identities on the lattice. Consider the following two supersymmetry transformations
\begin{align}
\notag \delta_1 \phi_i &= \psib_i \epsilon,            && \delta_2 \phi_i =  \bar{\epsilon} \psi_i, \\
\notag \delta_1 \psi_i &= -(D_{ik} \phi_k - W'_i) \epsilon,      && \delta_2 \psi_i = 0, \\
\delta_1 \psib_i &= 0,                           && \delta_2 \psib_i =  \bar{\epsilon} (D_{ik} \phi_k + W'_i),
\end{align}
where $\epsilon$ and $\bar{\epsilon}$ denote the infinitesimal Grassmann odd variables that generate supersymmetry. 

Let us take $\cO_{ij} = \phi_i \psib_j$. Upon using the transformation $\delta_2$ in the expression for the Ward identity we get
\bea
\langle \delta_2 \cO_{ij} \rangle = 0 ~~~~\implies~~~~ \langle (\phi_i \cdot \delta_2 \psib_j + \delta_2 \phi_i \cdot \psib_j ) \rangle = 0.
\eea

This leads to
\beq
\langle  \phi_i ( D_{jk} \phi_k + W'_j ) - \psib_j \psi_i \rangle = 0. 
\label{AA}
\eeq
  
We note that when we apply the transformation $\delta_1$ on the same observable $\phi_i \psib_j$ our result vanishes trivially. Taking $\widetilde{\cO}_{ij} = \phi_i \psi_j$ and applying the transformation $\delta_1$ on it, we get
\bea
\langle \delta_1 \widetilde{\cO}_{ij} \rangle = 0 ~~~~\implies~~~~ \langle (\phi_i \cdot \delta_1 \psi_j + \delta_1 \phi_i \cdot \psi_j ) \rangle &=& 0.
\eea
Upon simplification this becomes    
\beq
\label{A}
\langle \phi_i ( D _{jk} \phi_k - W'_j ) + \psib_i \psi_j \rangle = 0.
\eeq

Eqs. \eqref{AA} and \eqref{A} are our Ward identities. Note that the Ward identities are functions of sites $i$ and $j$. In the simulations we fix $i = 0$ and $j = n$ to monitor the Ward identities. We have
\bea
\label{w1}
w_1(n) &\equiv&  \langle  \phi_0 ( D_{nk} \phi_k + W'_n ) \rangle - \langle \psib_n \psi_0 \rangle, \\
\label{w2}
w_2(n) &\equiv&    \langle \phi_0 ( D_{nk} \phi_k - W'_n ) \rangle + \langle \psib_0 \psi_n \rangle.
\eea
If SUSY is preserved in the model these quantities should fluctuate around 0, at least in the middle region of the lattice, where the effects from higher excited states are the lowest. 

For supersymmetric quantum mechanics on the lattice we can use a simple scaling argument (see Ref. \cite{Catterall:2001fr}) to show that the expectation value of the total action
\beq
\langle S \rangle_{\rm exact} = \hf N_{\rm d.o.f.},
\eeq
where $N_{\rm d.o.f.}$ is the total number of degrees of freedom. For a lattice with $N_\tau$ sites, there are two degrees of freedom per site and thus there is a total of $2 N_\tau$ degrees of freedom in the lattice theory. Thus we have
\beq
\langle S \rangle_{\rm exact} = \hf N_{\rm d.o.f.} = \hf (2 N_\tau) = N_\tau.
\eeq

In our simulations we will use
\beq
\Delta S \equiv  \langle S \rangle  - \langle S \rangle_{\rm exact}  = \langle S \rangle - N_\tau
\eeq
as an indicator of SUSY breaking. 

To probe SUSY breaking we can also study the expectation value of the first derivative of the superpotential. This quantity is related to the auxiliary field $B$ through the relation $B = - W'$. We have $\langle W' \rangle = 0$ when SUSY is preserved and non-zero otherwise. (See the work by Kuroki and Sugino \cite{Kuroki:2009yg} for more details.) 

We will use a relatively more efficient algorithm, known as the Hybrid Monte Carlo (HMC) algorithm for the simulations. As part of this algorithm we take $p$ and $\pi$ as the momenta conjugate to the fields $\phi$ and $\chi$, respectively. Then the Hamiltonian of the system takes the form 
\beq
H = \hf \sum_i (p_i^2 + \pi_i^2) + S_B + S_F.
\eeq
This Hamiltonian is then evolved using a discretized version of Hamilton's equations, in a fictitious time $\tau$, using a small step size $\epsilon$. At the end of the evolution we get a new Hamiltonian $H'$. We then accept or reject $H'$ using the Metropolis test. (See Refs. \cite{Sexton:1992nu, Joseph:2019zer} for more details on the algorithm.)

\section{Supersymmetric Anharmonic Oscillator}
\label{sec:Supersymmetric Anharmonic Oscillator}

Let us consider the superpotential
\beq
W(\phi) = \hf m \phi^2 + \qtr g \phi^4,
\eeq
where $m$ is the mass and $g$ is the coupling constant. In Ref. \cite{Catterall:2000rv} it was concluded, using Monte Carlo simulations, that SUSY was preserved in this model. As a cross-check of our simulation code, we will reproduce the results given there.
  
The superpotential at a lattice site $i$ takes the following form
\bea
W'_i &=& \sum_{j = 0}^{N_\tau - 1} K_{ij} \phi_j + g \phi_i^3 \nn \\
&=& m \phi_i + \phi_i - \hf \left( \phi_{i-1} + \phi_{i+1} \right) + g \phi_i^3.
\eea

The fermionic matrix, introduced in Eq. \eqref{eq:ferm_matrix}, becomes
\bea
M_{ij} &=& \left(1 + m + 3 g \phi_i^2 \right) \delta_{ij} - \delta_{i, j+1}.
\eea

In the simulations we used the following dimensionless variables: $m = m_{\rm phys}a$, $g = g_{\rm phys}a^2$, and $\phi = \phi_{\rm phys} a^{-1/2}$. We performed the simulations for $N_\tau = 16, 32$, and $48$ with inverse temperature $\beta = 0.5, 1.0, 1.5$, and $2.0$. 

\begin{figure*}[htp]
\centering
\subfloat[$\langle W' \rangle$ per site against $N_\tau$ for various values of $\beta$.]{\includegraphics[width=7.5cm]{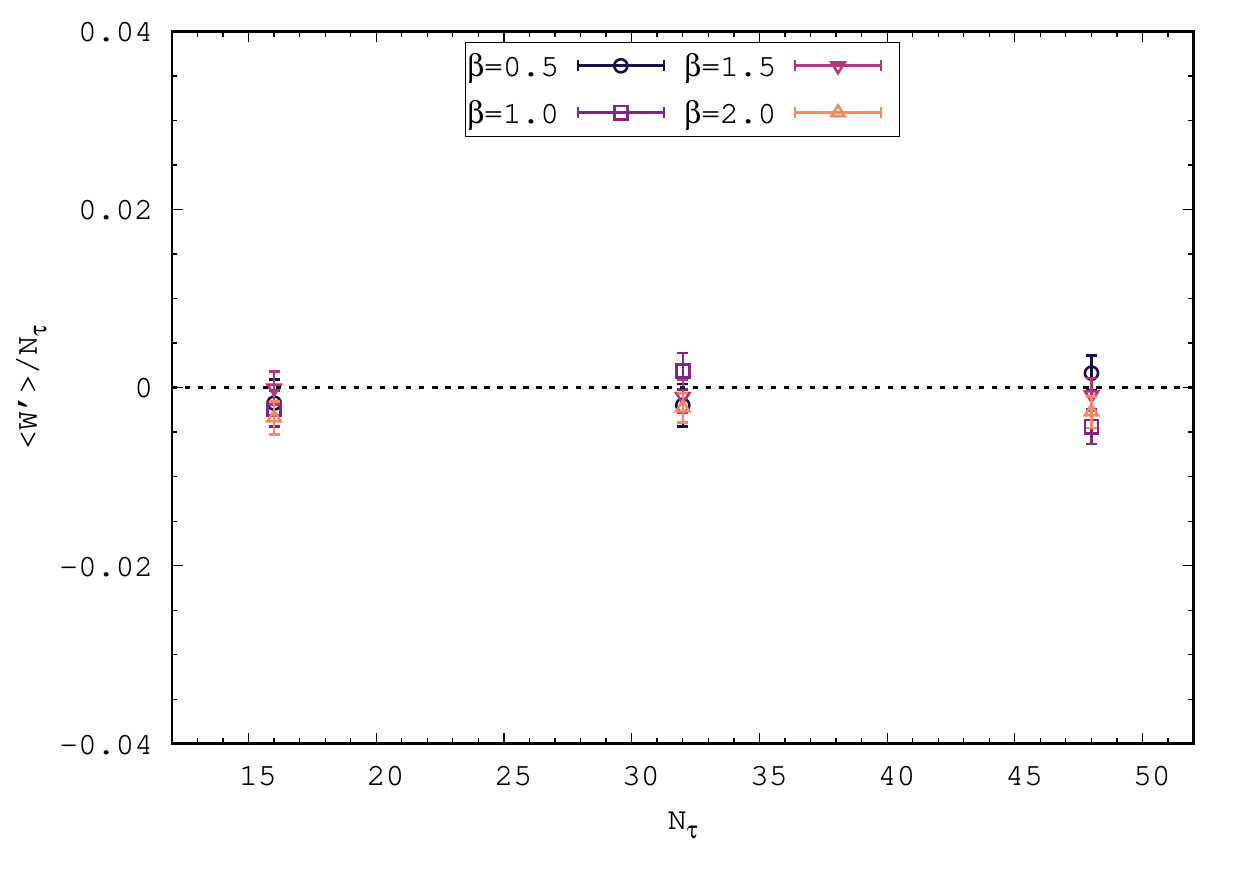}} $~~~$ \subfloat[$\Delta S$ per site against $N_\tau$ for various values of $\beta$.]{\includegraphics[width=7.5cm]{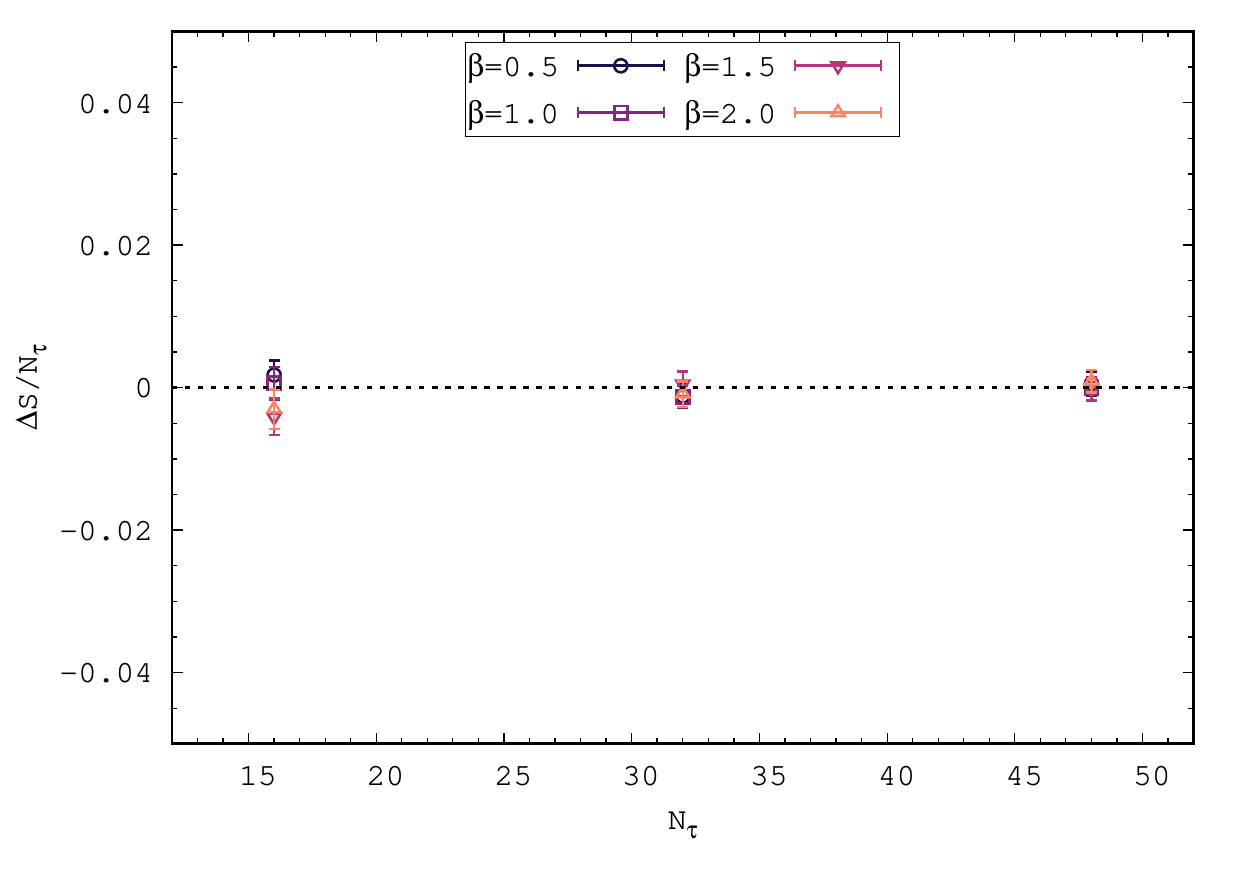}}

\caption{Supersymmetric anharmonic oscillator. (Left) Expectation value of $\langle W' \rangle$ per site against $N_\tau$ for various $\beta$ values. (Right) $\Delta S$ per site against $N_\tau$ for various $\beta$ values.}
\label{fig:act-deg-4_fig:w_p}

\end{figure*}

In Fig. \ref{fig:act-deg-4_fig:w_p} (left) $\langle W' \rangle$ per site against $N_\tau$ is shown for various $\beta$ values. The data fluctuate around zero within error bars for all $\beta$ values indicating that the model has intact SUSY. In Fig. \ref{fig:act-deg-4_fig:w_p} (right) we show the plot of $\Delta S$ per site against $N_\tau$ for various $\beta$ values. Here also the data fluctuate around zero within error bars for all $\beta$ values suggesting that SUSY is preserved in the model. 

\begin{figure*}[htp]
\centering
\subfloat[Ward identity given by Eq. \eqref{w1}.]{\includegraphics[width=7cm]{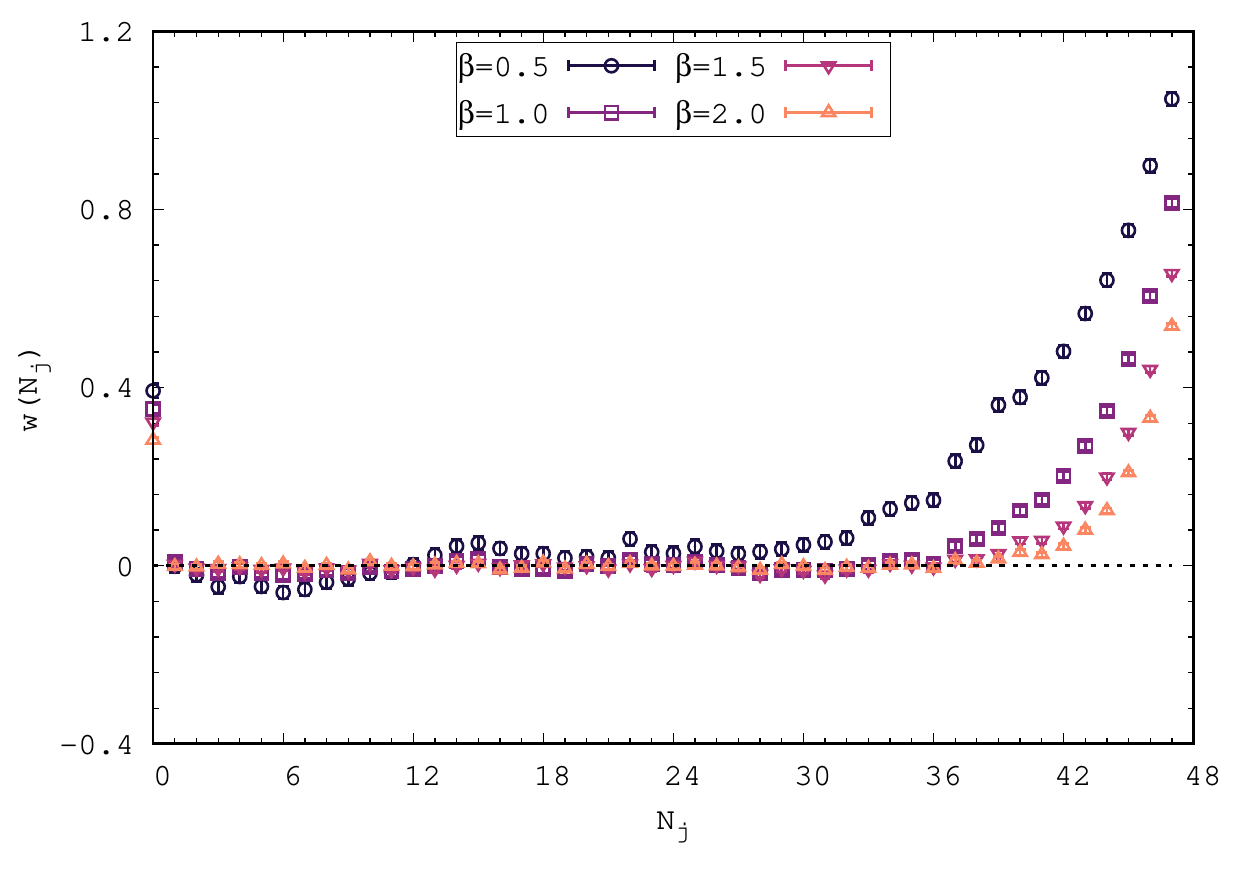}} $~~~$ \subfloat[Ward identity given by Eq. \eqref{w2}.]{\includegraphics[width=7cm]{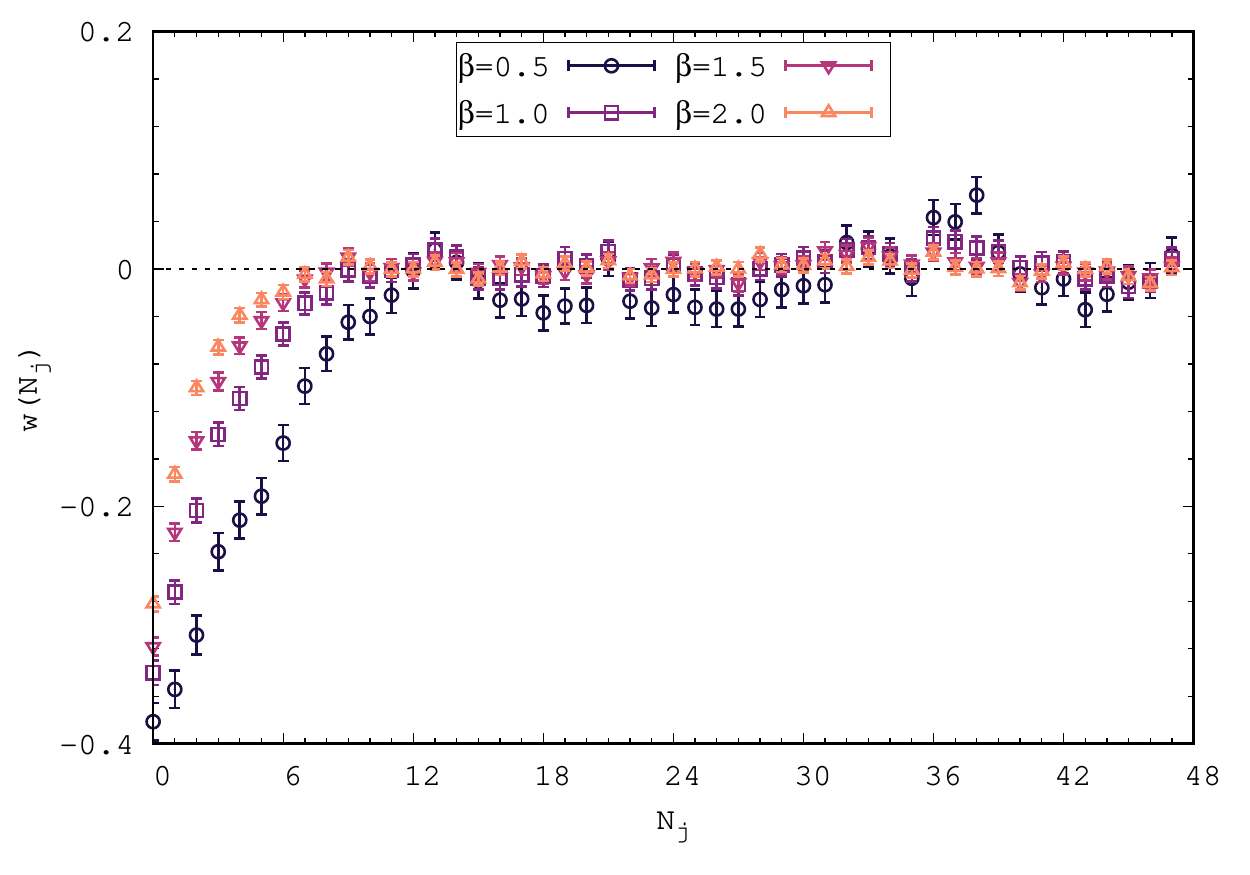}}

\caption{Supersymmetric anharmonic oscillator. Ward identities for different $\beta$ values on a lattice with $N_\tau = 48$. As $\beta$ is increased (temperature is decreased) the data approach closer to zero in the middle region of the lattice, as expected for a theory with intact SUSY.}
	\label{fig:ward_degree_4}
\end{figure*}

In Fig. \ref{fig:ward_degree_4} we show the simulation results for Ward identities. They show small fluctuations around zero in the middle regions of the lattice. The strength of the fluctuations tends to reduce as $\beta$ is increased (temperature is reduced). Overall, the data suggest that SUSY is preserved in this model. 

Taking into account the combined results shown in Figs. \ref{fig:act-deg-4_fig:w_p} and \ref{fig:ward_degree_4} we conclude that SUSY is preserved in this model.

\section{Model with Odd-Degree Superpotential}
\label{sec:Model_with_Odd-Degree_Superpotential}

In this section we consider a model with degree-five superpotential. According to Ref. \cite{Kanamori:2007yx} SUSY is dynamically broken in this model. 

Our lattice prescription gives the following form for the derivative of the superpotential
\bea
\label{n1}
W'_i &=& \sum_{j = 0}^{N_\tau - 1} K_{ij} \phi_j + g \phi_i^4 \nn \\
&=& m \phi_i + \phi_i - \hf \left( \phi_{i-1} + \phi_{i+1} \right) + g \phi_i^4.
\eea

The dimensionless parameters are $m = m_{\rm phys} a$, $g = g_{\rm phys} a^{5/2}$ and $\phi = \phi_{\rm phys} a^{-1/2}$. We simulated this model with the parameters $m = 10$ and $g = 100$. 

\begin{figure*}[htp]

\subfloat[$\langle W' \rangle $ per site against $N_\tau$.]{\includegraphics[width=7.5cm]{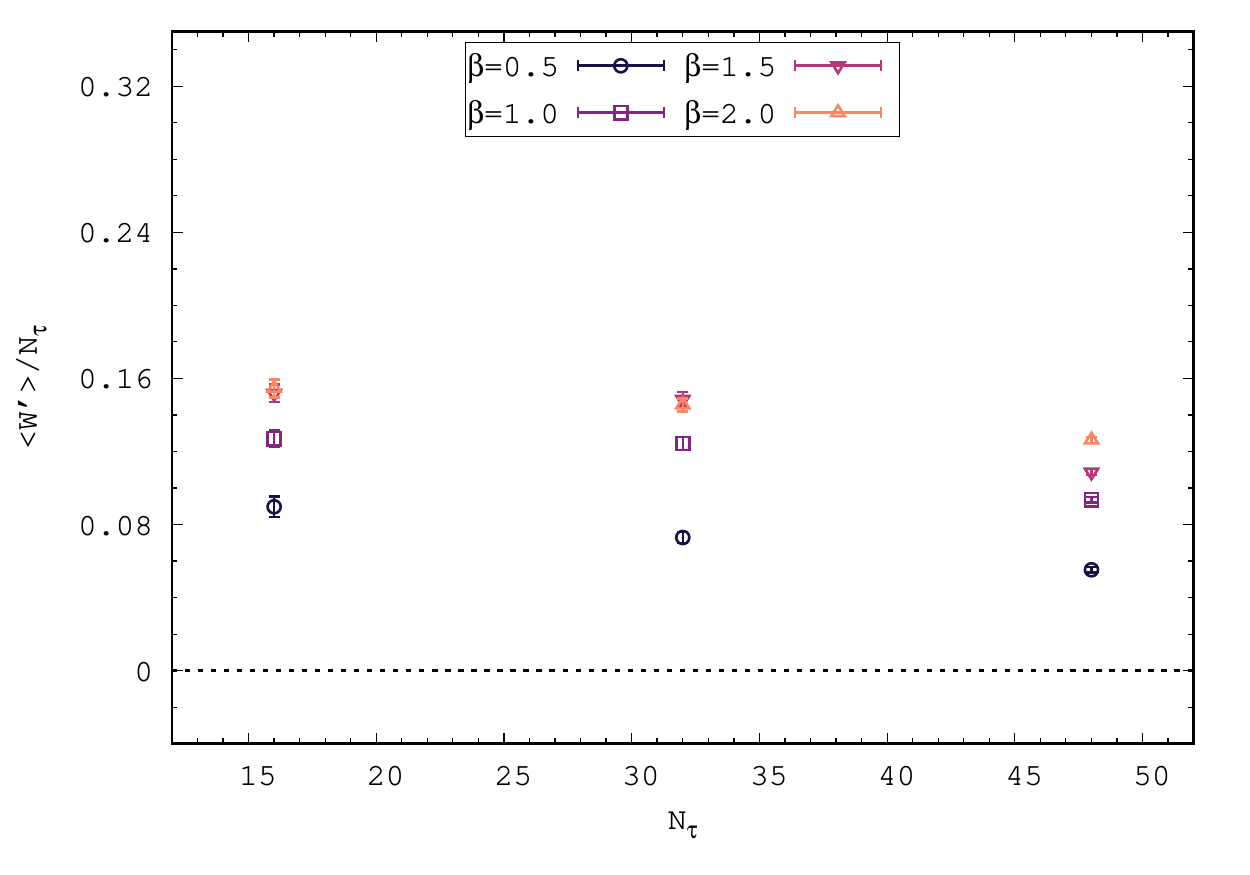}} $~~~$ \subfloat[$\Delta S$ per site against $N_\tau$ for various values of $\beta$.]{\includegraphics[width=7.5cm]{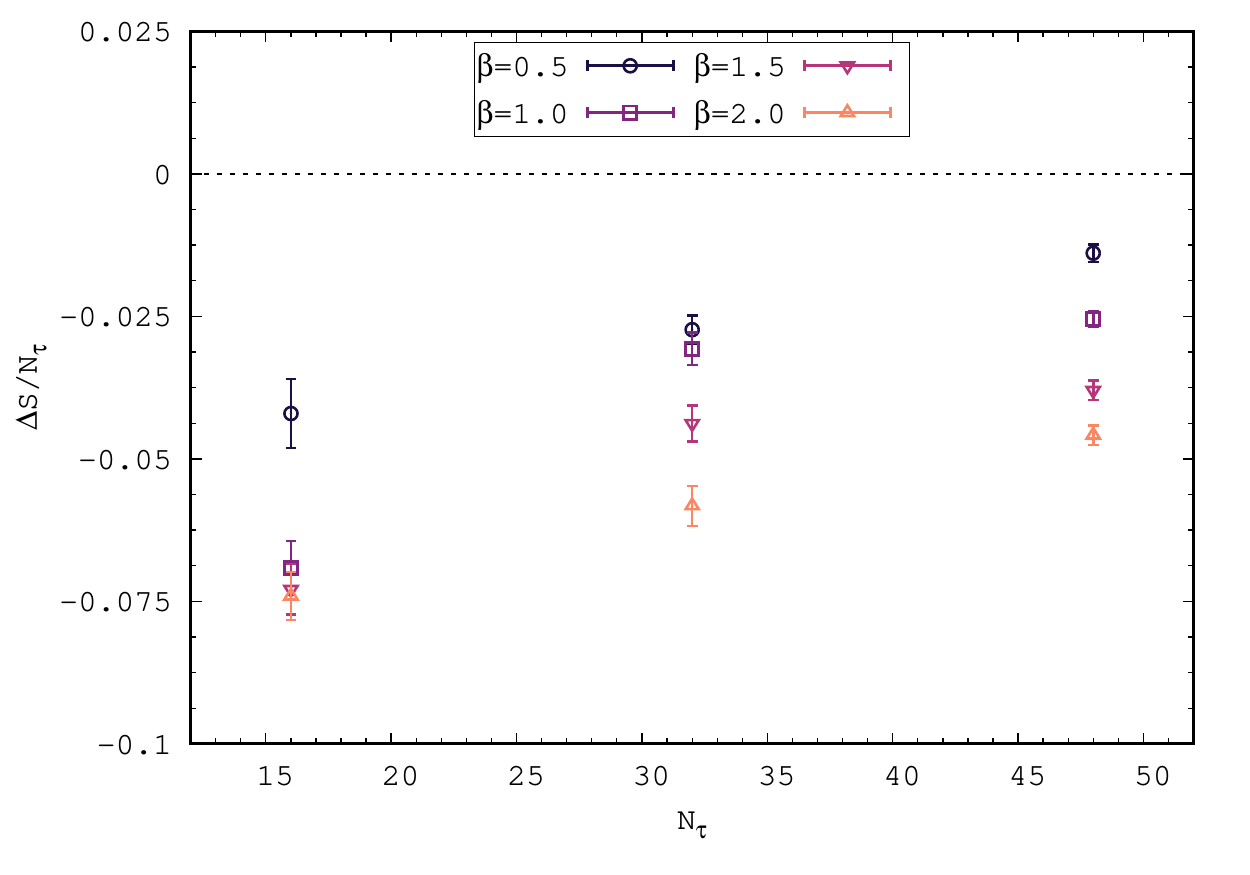}} 

\caption{Model with degree-five superpotential. (Left) Expectation value of $\langle W' \rangle$ per site against $N_\tau$. (Right) $\Delta S$ per site against $N_\tau$ for various $\beta$ values.}
\label{fig:act-deg-5_fig:w_prime_deg-5}
\end{figure*}

In Fig. \ref{fig:act-deg-5_fig:w_prime_deg-5} (left) $\langle W' \rangle$ per site against $N_\tau$ is shown for various $\beta$ values. It is non-vanishing for all $\beta$ values and deviates away from zero as $\beta$ (temperature) is increased (decreased). In Fig. \ref{fig:act-deg-5_fig:w_prime_deg-5} (right) we show the plot of $\Delta S$ per site against $N_\tau$ for various $\beta$ values. The data do not fluctuate around zero. As $\beta$ is increased $\Delta S$ increases. Thus we conclude that SUSY is broken in this model.

\begin{figure*}[htp]
\centering
\subfloat[Ward identity given by Eq. \eqref{w1}.]{\includegraphics[width=7.5cm]{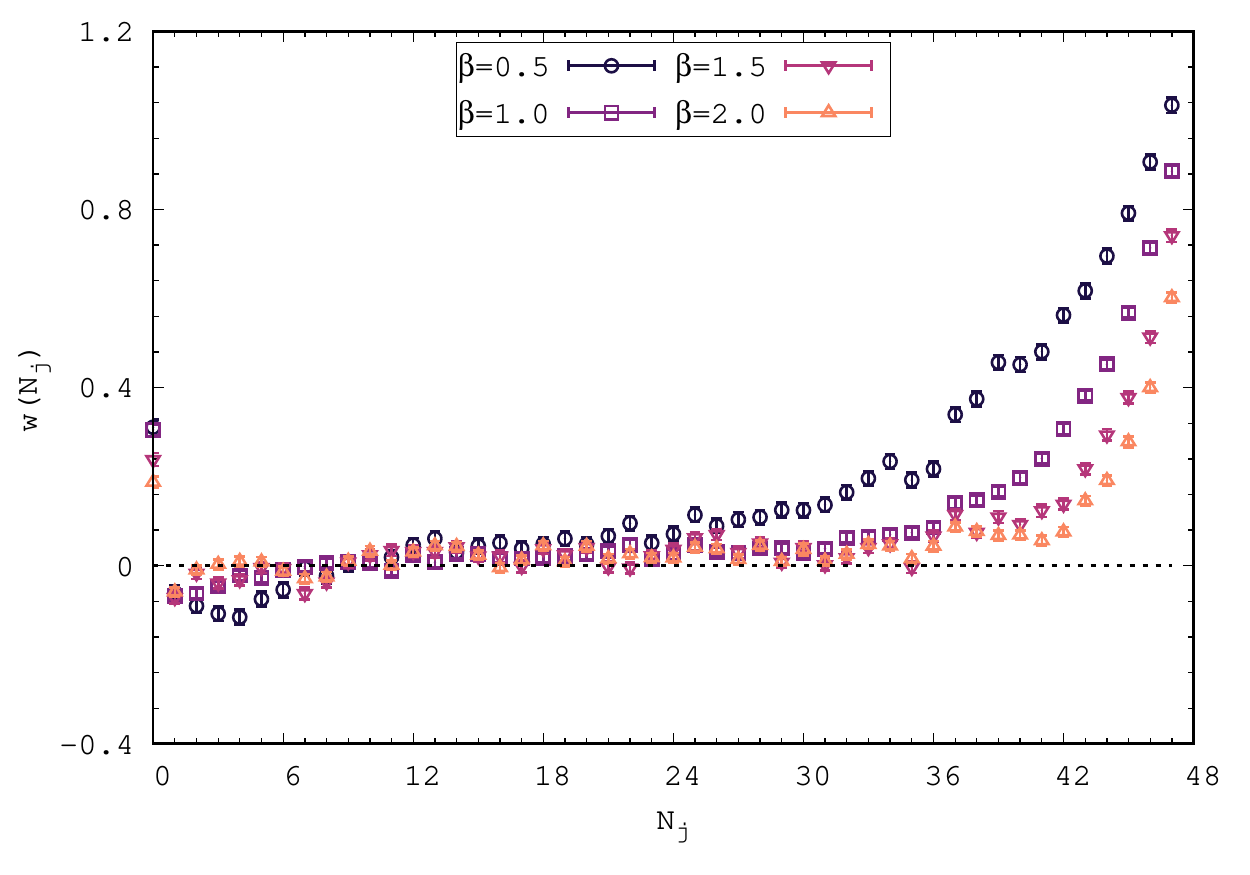}} $~~~$ \subfloat[Ward identity given by Eq. \eqref{w2}.]{\includegraphics[width=7.5cm]{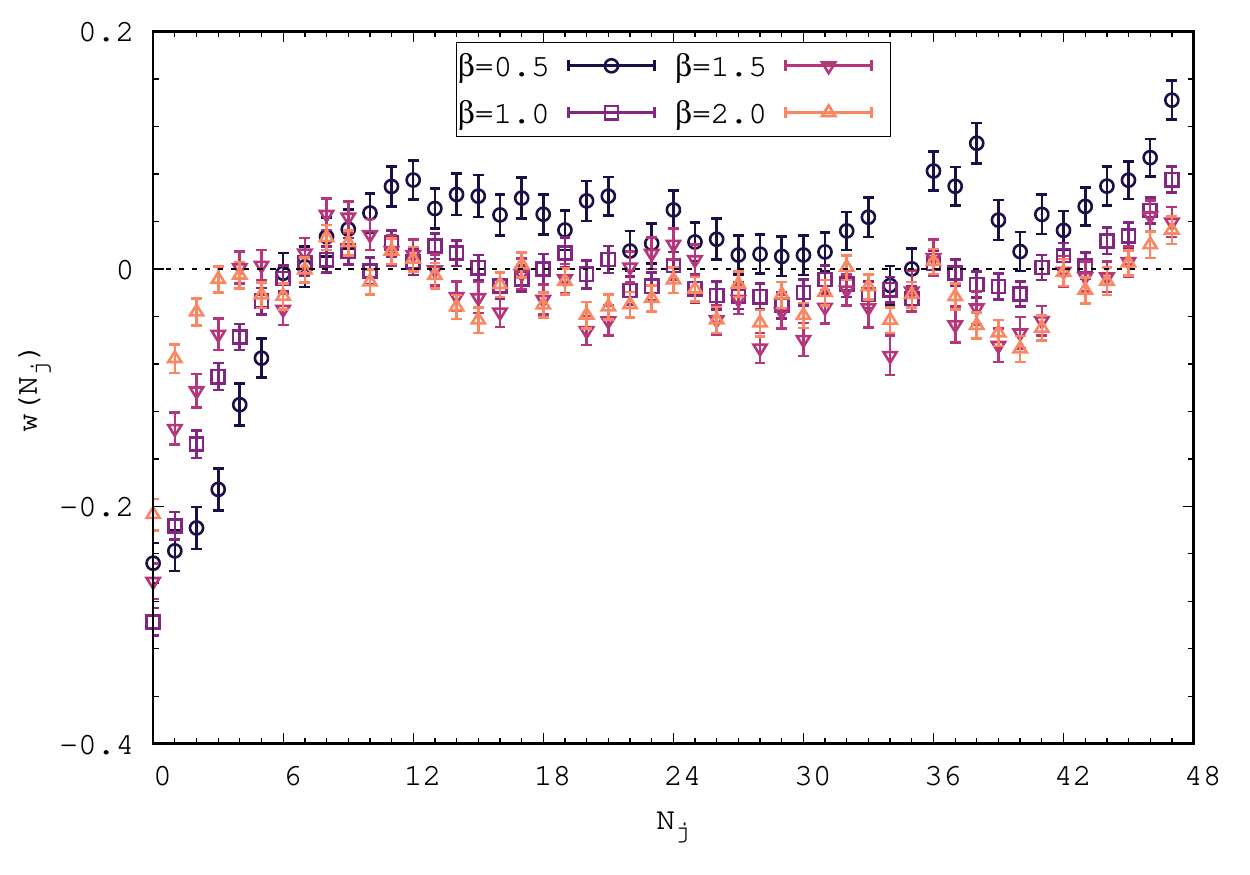}}
	
	\caption{Model with degree-five superpotential. Ward identities for the model with different $\beta$ values on a lattice with $N_\tau = 48$.}
	\label{fig:ward_degree_5}
\end{figure*}

The simulation results for Ward identities are shown in Fig. \ref{fig:ward_degree_5}. On the left panel we show the Ward identity Eq. \eqref{w1}. We see that the fluctuations are reducing in the middle region of the lattice as $\beta$ is increased. It is not possible to conclude whether SUSY is broken or not from this data alone. On the right panel the Ward identity Eq. \eqref{w2} exhibits large fluctuations around zero in the middle region of the lattice and it is not diminishing as $\beta$ is increased (temperature is decreased), thus suggesting that SUSY is broken in the model.

Taking into consideration the overall trend from Figs. \ref{fig:act-deg-5_fig:w_prime_deg-5} and \ref{fig:ward_degree_5} we conclude that SUSY is broken in the model with degree-five superpotential.

\section{Model with Scarf I Superpotential}
\label{sec:Scarf_I_Superpotential}

In this section, we show the simulation results for the model with a particular type of shape invariant potential, the Scarf-I superpotential \cite{Dutt:1986va, Cooper:1994eh}, and investigate the absence or presence of non-perturbative SUSY breaking.

The potential has the form
\beq
W'(\phi) = A~\text{tan$\left(\alpha \phi \right)$}-B~\text{sec$\left(\alpha \phi \right)$}, ~~~~ -\frac{\pi}{2}\le \alpha \phi  \le \frac{\pi}{2},
\eeq 
where $A >  B \ge 0$ and $\alpha > 0$. 

We will focus on the case $B = 0$. Then
\beq
\label{eqn:lat-scarf-pot}
W'(\phi) = \lambda \alpha ~\text{tan$\left(\alpha \phi \right)$}, ~~~~ -\frac{\pi}{2}\le \alpha \phi  \le \frac{\pi}{2}.
\eeq
The parameter $\alpha$ has the dimension of square root of energy and $\lambda$ is a dimensionless coupling.

This particular model was studied recently by Kadoh and Nakayama, in Ref. \cite{Kadoh:2018ele}, using a direct computational approach, based on transfer matrix formalism. It was shown there that SUSY was preserved in this model. 

With this potential, we use the dimensionless parameters, $\alpha$ = $\alpha_{\rm phys}a^{1/2}$, $\lambda$ = $\lambda_{\rm phys}$ and $\phi$ = $\phi_{\rm phys}a^{-1/2}$ on the lattice. 

At a lattice site $k$ the derivative of the superpotential has the form $W_k^{'} = \lambda \alpha \tan(\alpha \phi_k)$. 

\begin{figure*}[htp]
\centering
\subfloat[$\langle W' \rangle $ per site vs $N_\tau$.]{\includegraphics[width=7cm]{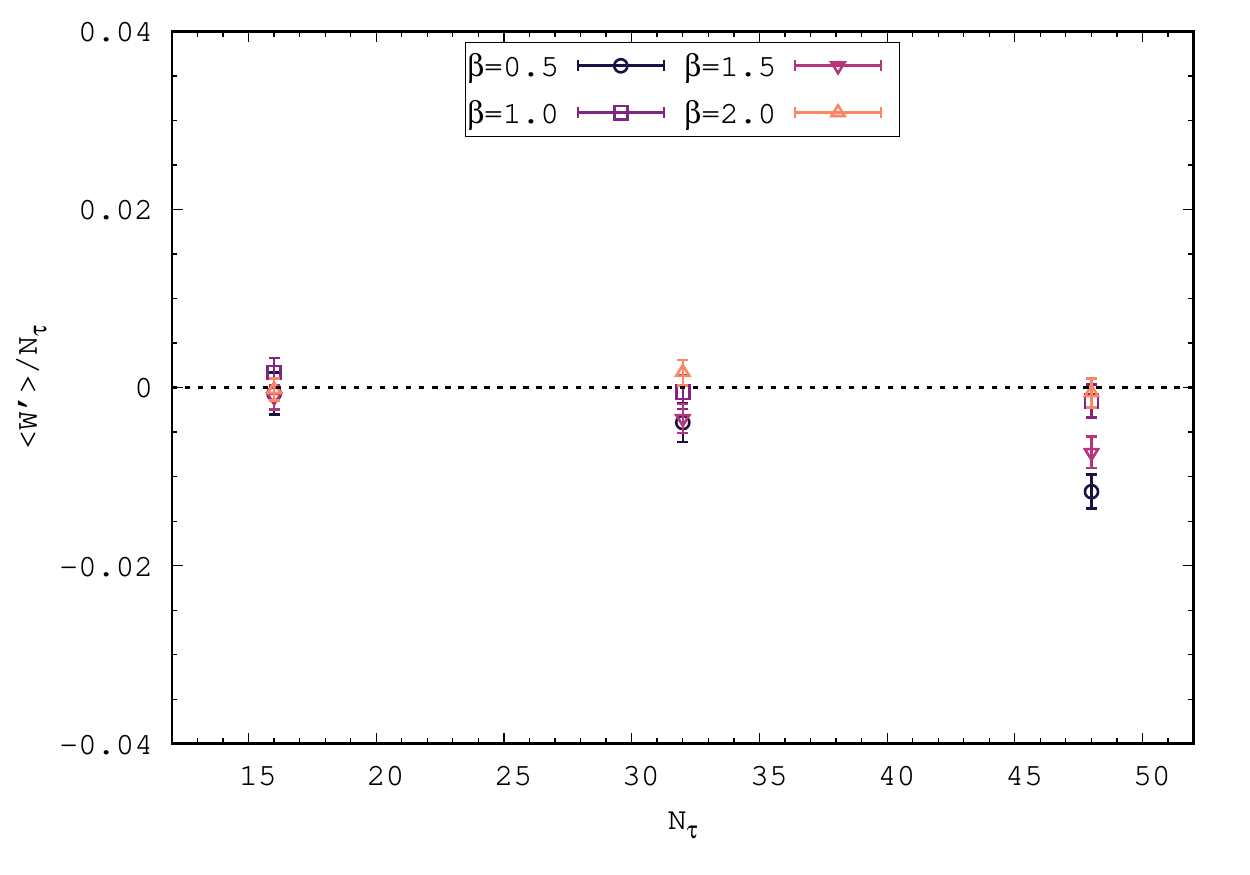}} $~~~$ \subfloat[$\Delta S$ per site for various values of $\beta$.]{\includegraphics[width=7cm]{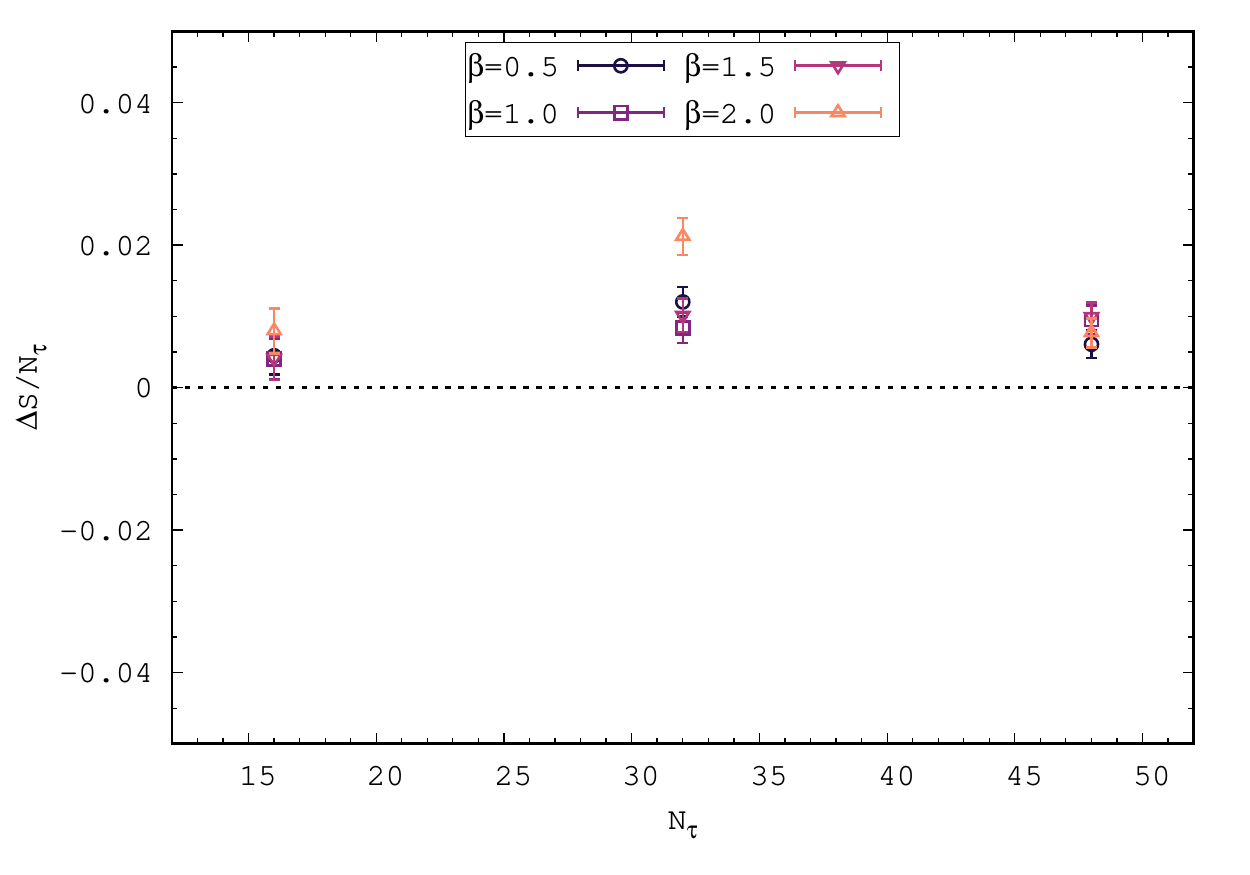}} 
	
	\caption{Model with Scarf I superpotential. (Left) Expectation value of $\langle W' \rangle$ per site against $N_\tau$ for various $\beta$ values. It approaches zero as $\beta$ is increased (temperature is decreased) agreeing with what is expected for a model with intact SUSY. (Right) $\Delta S$ per site against $N_\tau$ for various $\beta$ values. We do not see a consistent trend suggesting intact SUSY from this observable.}
	\label{fig:action_per_site_scarf_fig:W_prime_scarf}
\end{figure*}

In Fig. \ref{fig:action_per_site_scarf_fig:W_prime_scarf} (left) we show $W'$ per site against $N_\tau$. As $\beta$ is increased (temperature is decreased) this observable approaches zero (data points with triangle symbols) suggesting that SUSY is preserved in this model. However, Fig. \ref{fig:action_per_site_scarf_fig:W_prime_scarf} (right) shows $\Delta S$ per site against $N_\tau$ for various $\beta$ values. Although all data points are close to zero, they do not point towards a consistent trend suggesting an intact SUSY.

\begin{figure}[htp]
\centering
	
	\subfloat[Ward identity given by Eq. \eqref{w1}.]{\includegraphics[width=7cm]{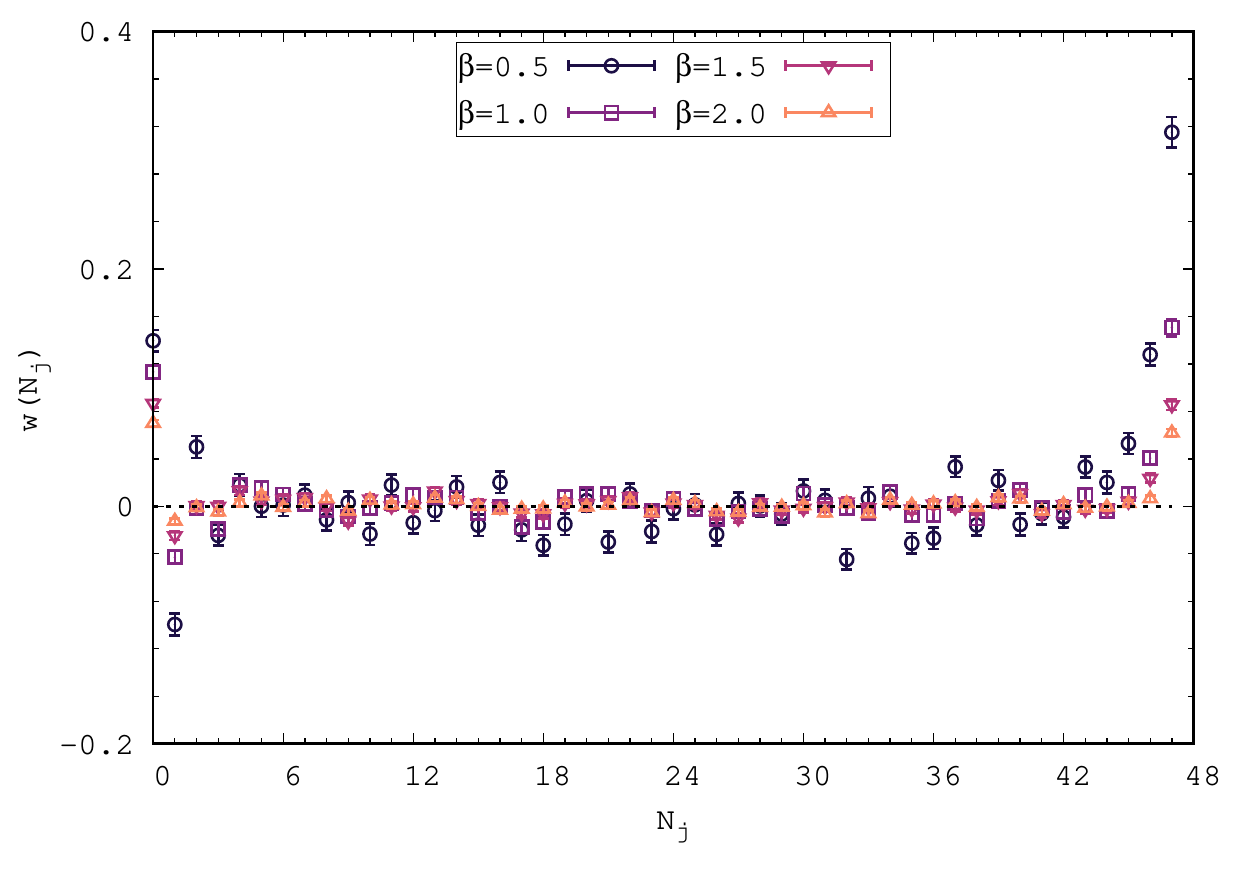}}
	\subfloat[Ward identity given by Eq. \eqref{w2}.]{\includegraphics[width=7cm]{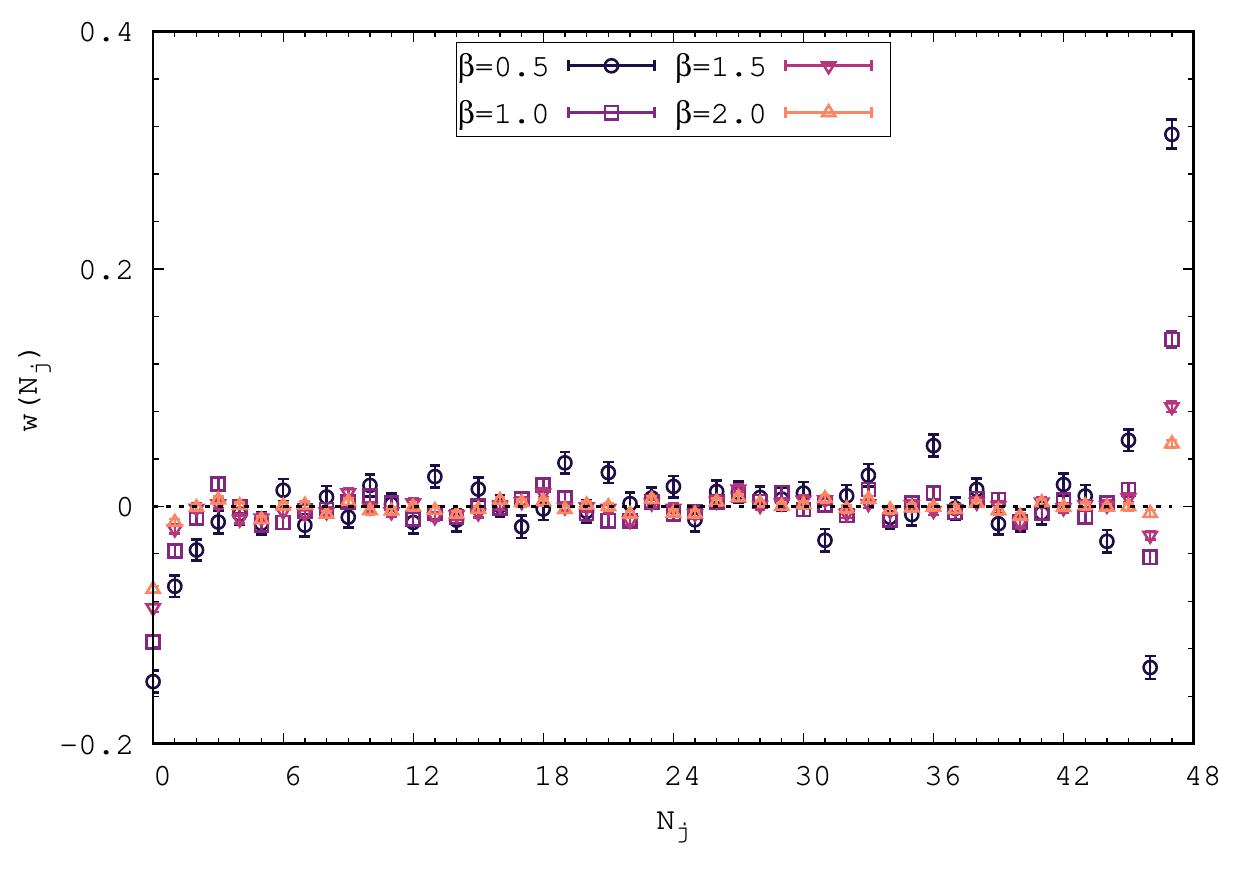}}
	\caption{Model with Scarf I potential. Ward identities for various $\beta$ values on a lattice with $N_\tau = 48$. The dimensionless parameters used are $\lambda = 10$ and $\alpha_{\rm phys} = \sqrt{60}$. In the middle region of the lattice, the Ward identities approach closer to zero as $\beta$ is increased (temperature is decreased) indicating unbroken SUSY in this model.}
	\label{Fig:ward}
\end{figure}

The Ward identities for various $\beta$ values are shown in Fig. \ref{Fig:ward} for a lattice with $N_\tau = 48$. We see that in the middle region of the lattice, the Ward identities approach closer to zero as $\beta$ is increased (temperature is decreased) suggesting that SUSY is intact in this model.

Taking into account the overall trend of the data from Figs. \ref{fig:action_per_site_scarf_fig:W_prime_scarf} and \ref{Fig:ward} we can conclude that SUSY is preserved in the model with Scarf I superpotential.

\section{Models Exhibiting $PT$-Symmetry}
\label{sec:SUSYQM_with_PT-Symmetry}

Another interesting class of models we investigated on the lattice is the quantum mechanics with certain type of $PT$-invariant superpotentials. Here, $P$ and $T$ denote the parity symmetry and time reversal invariance, respectively. One of the motivations for considering $PT$-symmetric theories is the following. It is possible to obtain a real and bounded spectrum if we impose $PT$-symmetric boundary conditions on the functional-integral representation of the four-dimensional $- \lambda \phi^4$ theory \cite{Bender:1999ek}. In addition, the theory becomes perturbatively renormalizable and asymptotically free. These properties suggest that a $- \lambda \phi^4$ quantum field theory might be useful in describing the Higgs sector of the Standard Model. It could also play a vital role in the supersymmetric versions of the Standard Model. Our investigations of the one-dimensional model can be considered as a first step towards explorations in this direction. 

In Ref. \cite{Bender:1997ps} it was shown that a two-dimensional supersymmetric field theory, with a $PT$-symmetric superpotential of the form 
\beq
W(\phi) = \frac{-g}{2 + \delta} \left( i \phi \right)^{2 + \delta},
\eeq
with $\delta$ a positive parameter, exhibited intact SUSY. A recent study using complex Langevin dynamics showed that SUSY is not dynamically broken in these models in zero and one dimensions \cite{Joseph:2019sof, Joseph:2020gdh}. Since Monte Carlo is reliable only when the action is real, we simulated a subset of these $PT$-symmetric potentials, which have real actions. For a full analysis of the model with a general $\delta$ parameter, complex Langevin dynamics or any other compatible method should be used.

At a lattice site $k$, the $PT$-invariant superpotential has the following form
\beq
W_k = \frac{-g}{2 + \delta} \left( i \phi_k \right)^{2 + \delta}.
\eeq

The derivative of the superpotential is
\beq
\label{W'}
W'_k = \sum_j K_{kj} \phi_j - i g \left( i \phi_k \right)^{1 + \delta}.
\eeq

Although we have massless theories as the continuum cousins, we need to introduce Wilson-type mass terms in the simulations, as shown in Eq. \eqref{W'}. We perform simulations for various mass values and then take the limit $m \rightarrow 0$. Since the action needs to be real for Monte Carlo simulations to be reliable, we have investigated only models with $\delta = 0, 2$, and 4. 

The superpotentials take the following forms for these $\delta$ values 
\bea
\delta = 0: ~~~~ W'_k &=& \phi_k + m \phi_k - \hf \left( \phi_{k-1} + \phi_{k+1} \right) + g \phi_k, \\
\delta = 2: ~~~~ W'_k&=& \phi_k + m \phi_k - \hf \left( \phi_{k-1} + \phi_{k+1} \right) - g \phi_k^3, \\
\delta = 4: ~~~~ W'_k &=& \phi_k + m \phi_k - \hf \left( \phi_{k-1} + \phi_{k+1} \right) + g \phi_k^5.
\eea

\begin{figure*}[htp]
\centering	
\subfloat[$\langle W' \rangle $ per lattice site against $N_\tau$.]{\includegraphics[width=7cm]{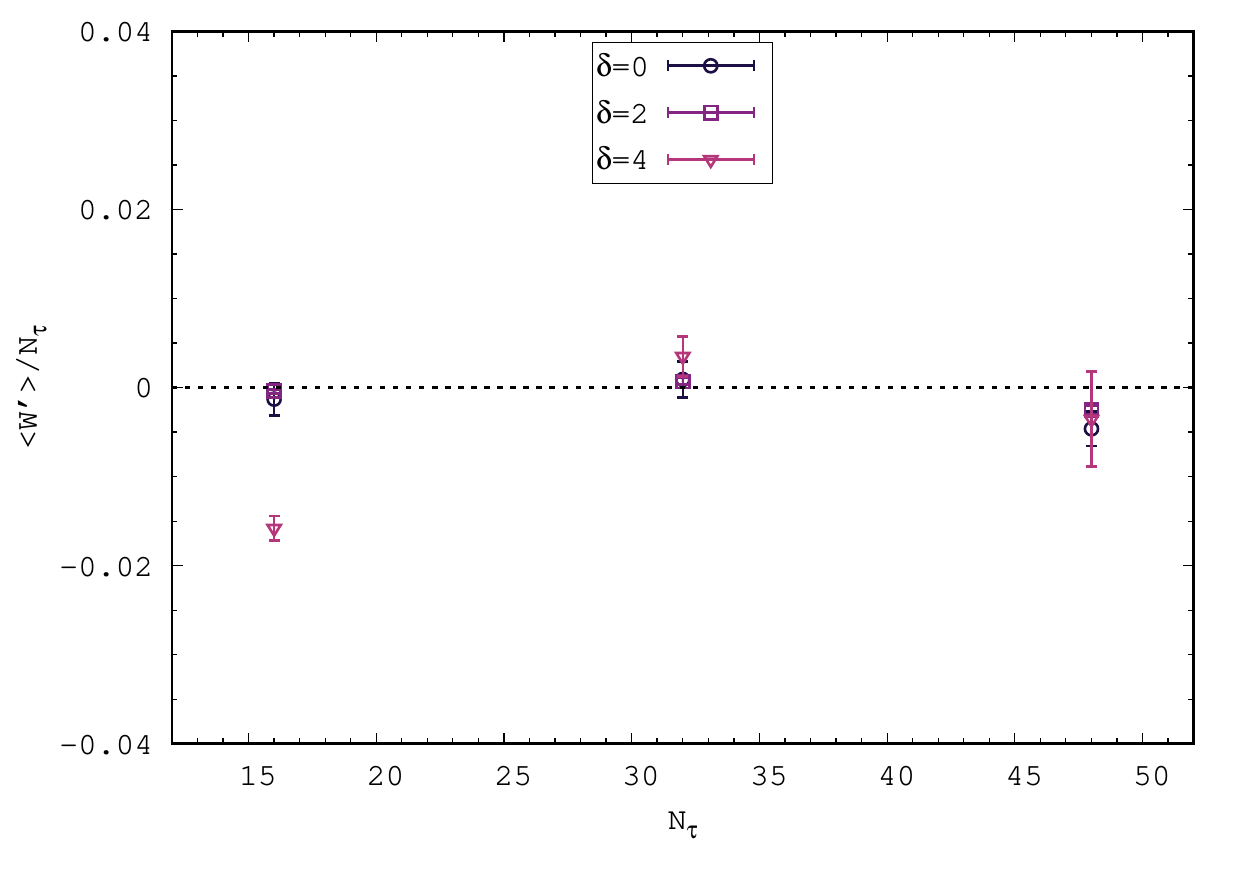}} $~~~$ \subfloat[$\Delta S$ per site against $N_\tau$.]{\includegraphics[width=7cm]{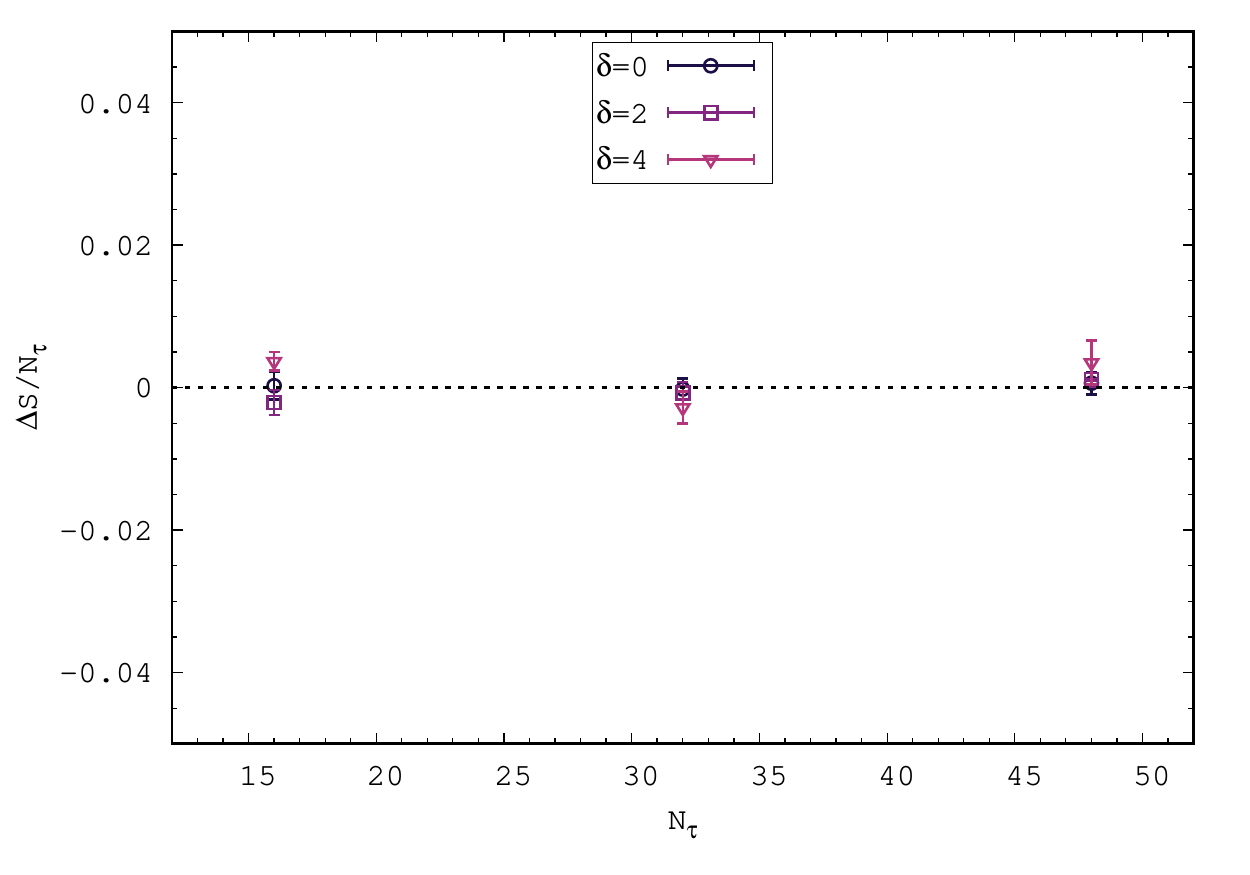}} 
	\caption{Model with $PT$-symmetric superpotentials. (Left) Expectation value of $\langle W' \rangle$ per site against $N_\tau$. (Right) $\Delta S$ per site against $N_\tau$.}
	\label{fig:action_per_site_pt_fig:W_prime_pt}
\end{figure*}

\begin{figure}[htp]
\centering
	
	\subfloat[Ward identity Eq. \eqref{w1} for $\delta = 0$.]{\includegraphics[width=7cm]{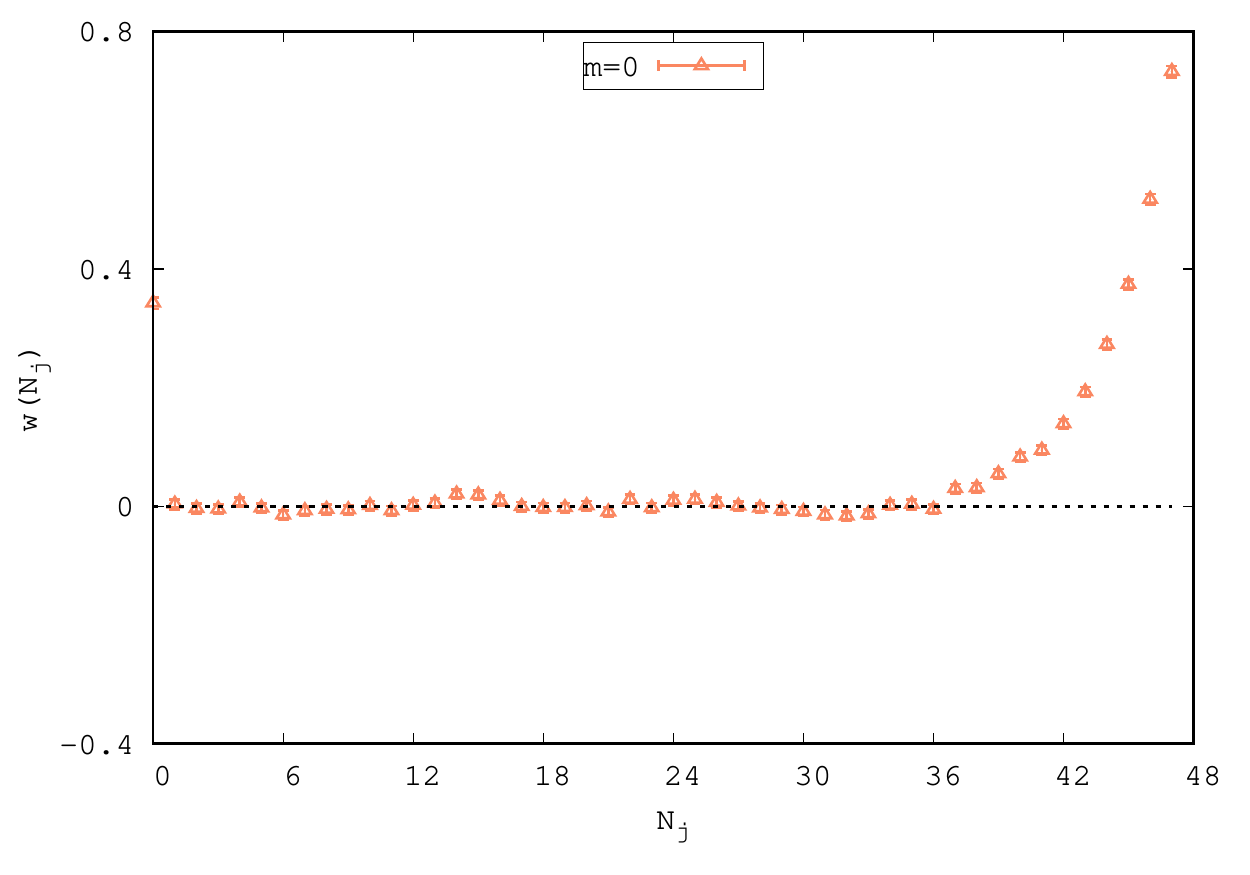}}
	\subfloat[Ward identity Eq. \eqref{w2} for $\delta = 0$.]{\includegraphics[width=7cm]{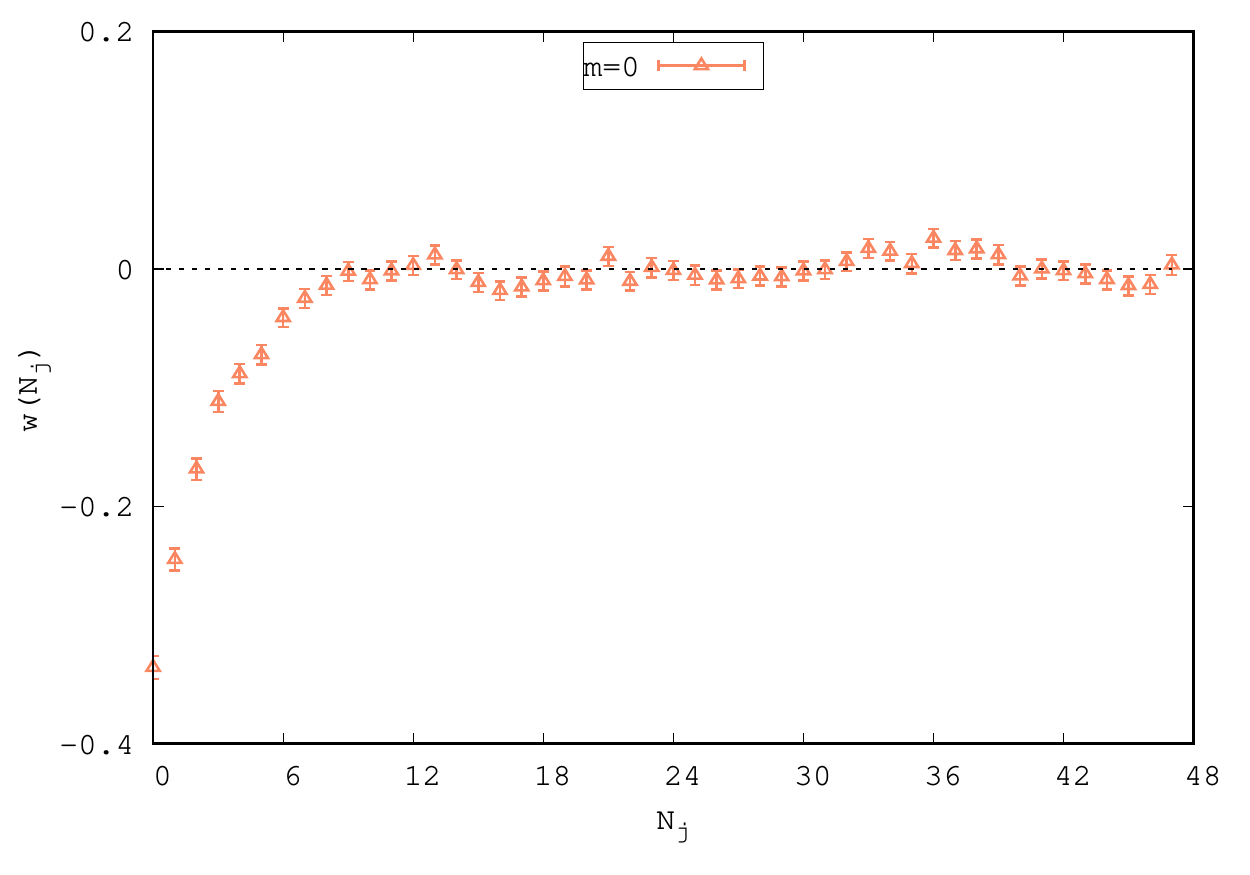}}
	
	\subfloat[Ward identity Eq. \eqref{w1} for $\delta = 2$.]{\includegraphics[width=7cm]{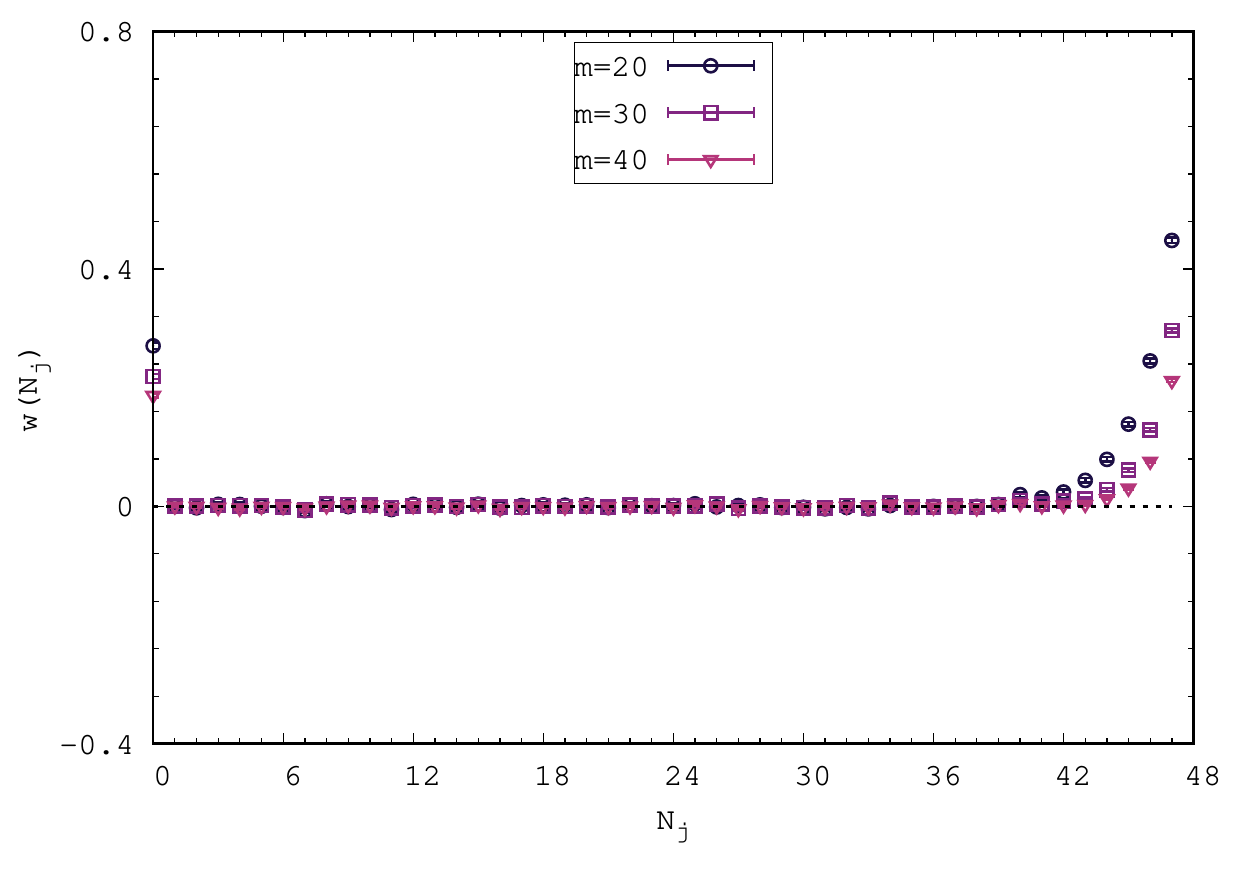}}
	\subfloat[Ward identity Eq. \eqref{w2} for $\delta = 2$.]{\includegraphics[width=7cm]{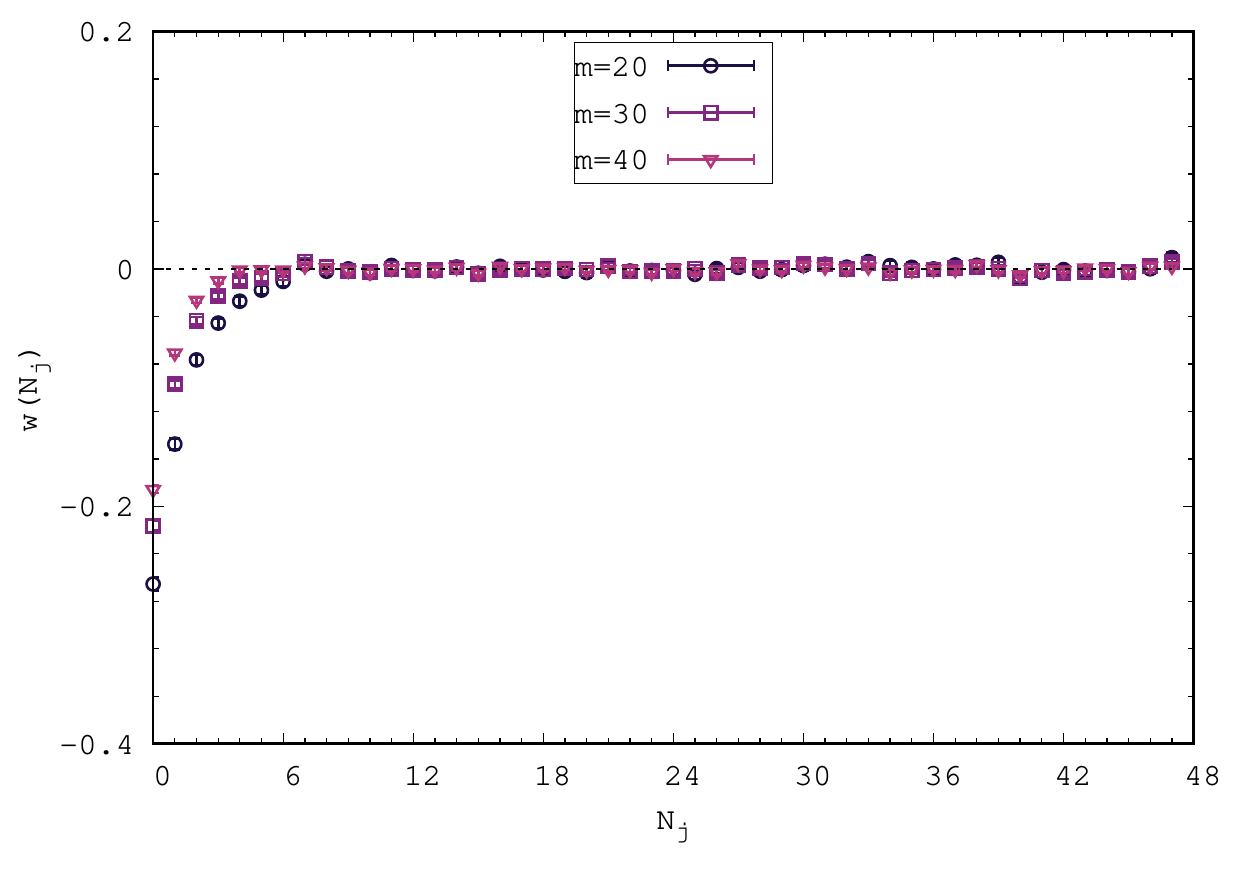}}
	
	\subfloat[Ward identity Eq. \eqref{w1} for $\delta = 4$.]{\includegraphics[width=7cm]{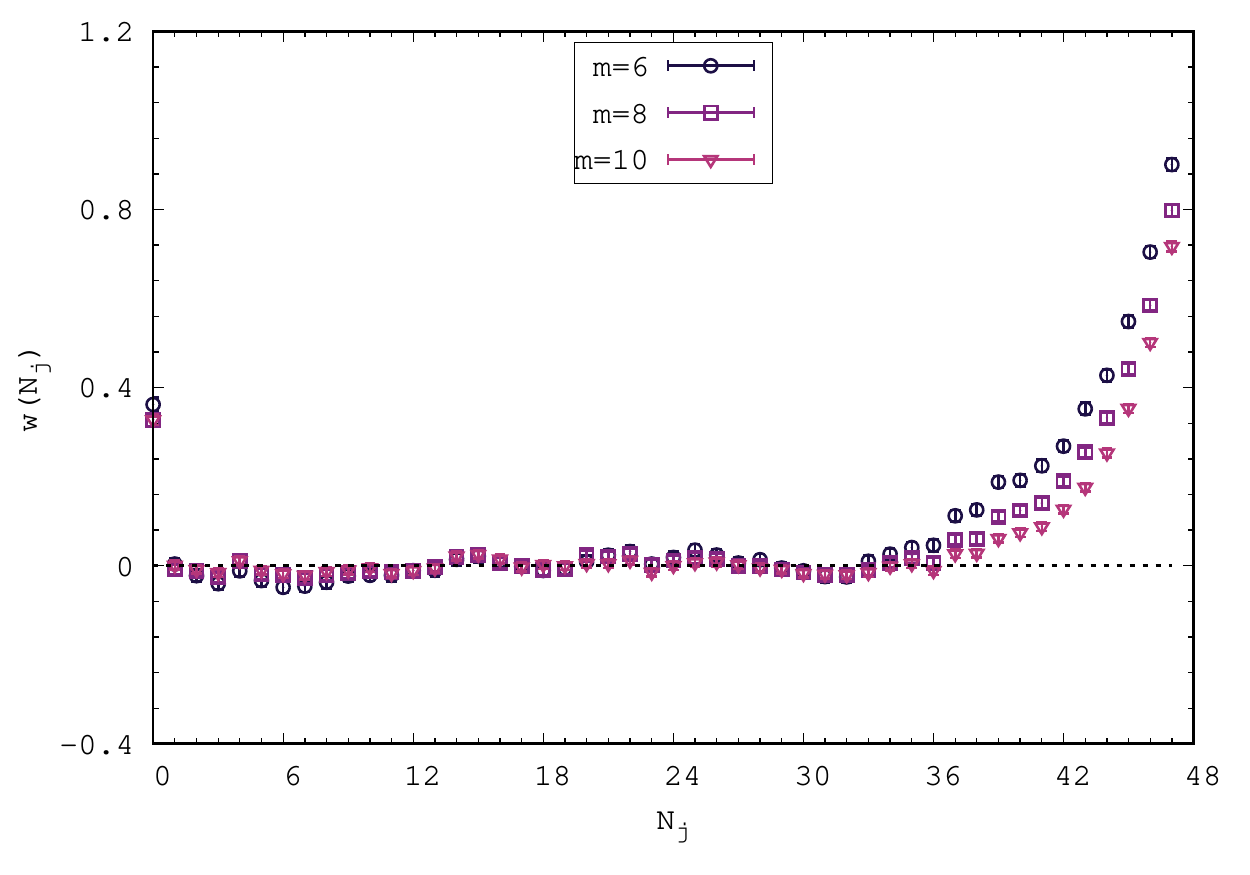}}
	\subfloat[Ward identity Eq. \eqref{w2} for $\delta = 4$.]{\includegraphics[width=7cm]{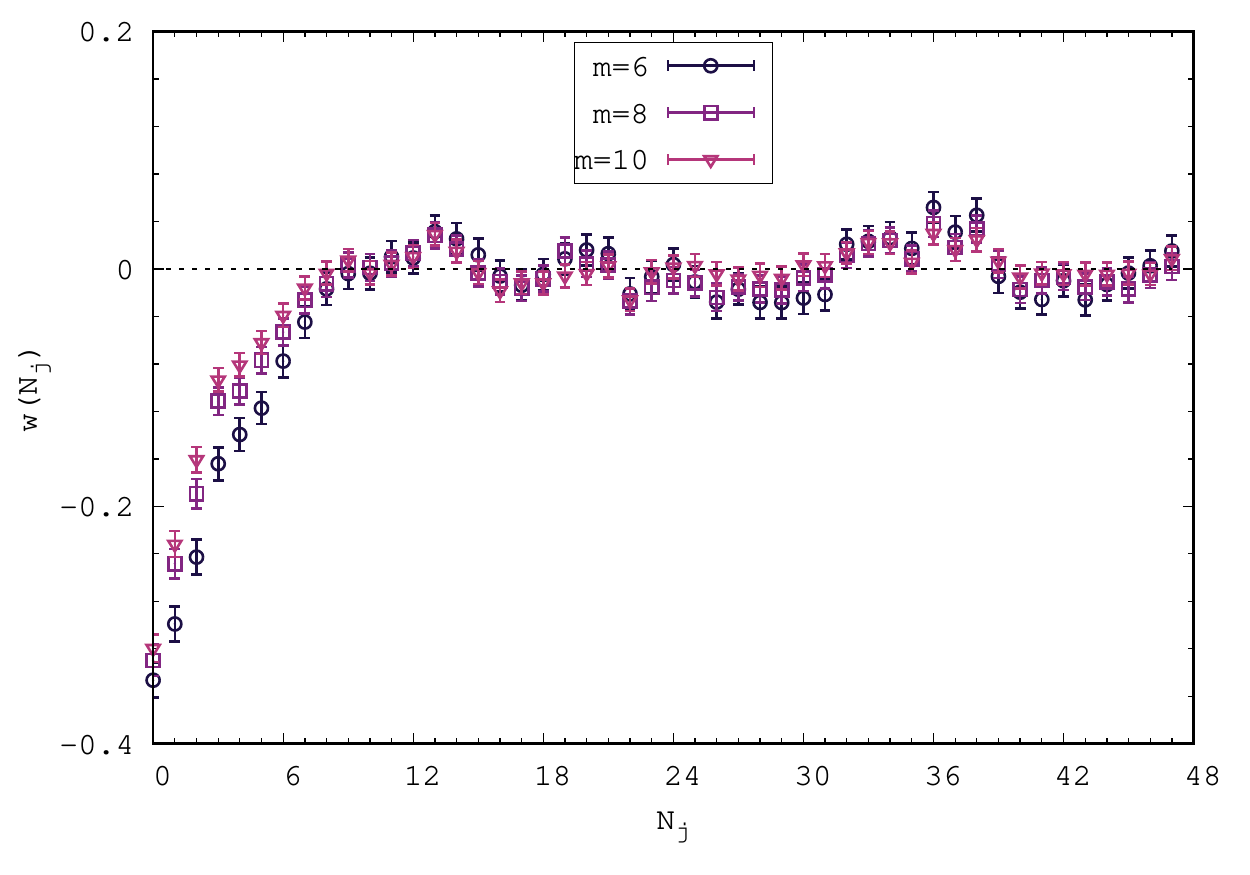}}
	\caption{Models with $PT$-symmetric superpotentials. Ward identities on a $N_\tau = 48$ lattice. We used $\beta = 2.0$ in the simulations.}
	\label{fig:ward_pt}
\end{figure}

In Fig. \ref{fig:action_per_site_pt_fig:W_prime_pt} (left) we show $W'$ per site against $N_\tau$ for various $\delta$ values. In Fig. \ref{fig:action_per_site_pt_fig:W_prime_pt} (right) we show $\Delta S$ per site against $N_\tau$ for various $\delta$ values. Both the plots are produced at $m = 0$. For $\delta = 0$, we were able to simulate at $m = 0$ value itself, and for $\delta = 2, 4$ we simulated with various $m$ values and then took the $m \rightarrow 0$ limit. The data in Fig. \ref{fig:action_per_site_pt_fig:W_prime_pt} (left) do not lie exactly on top of zero making it difficult to conclude straight away that SUSY is preserved in this model. The data in Fig. \ref{fig:action_per_site_pt_fig:W_prime_pt} (right) fluctuate around zero within error bars for all $\beta$ values suggesting that SUSY is preserved in this model. In order to get a better insight we also investigated the Ward identities for this model. The plots are shown in Fig. \ref{fig:ward_pt}. There, the Ward identities are plotted for all the mass parameter values used in the simulations. We used $m = 0$ for $\delta = 0$; $m = 20, 30, 40$ for $\delta = 2$; and $m = 6, 8, 10$ for $\delta = 4$ in the simulations. 

 In Fig. \ref{fig:ward_pt} the data fluctuate around zero in the middle regions of the lattice, for all the $\delta$ values, suggesting that SUSY is preserved in these models. Thus taking into consideration the overall behavior of the data in Figs. \ref{fig:action_per_site_pt_fig:W_prime_pt} and \ref{fig:ward_pt}, we conclude that SUSY is preserved in these models with $PT$-invariant potentials.

\section{Conclusions and Future Directions}
\label{sec:Conclusions}

In this paper, we have investigated non-perturbative SUSY breaking in quantum mechanics models with various types of superpotentials using lattice regularized path integrals. We employed Hybrid Monte Carlo (HMC) algorithm to perform the field updates on the lattice. After reproducing the existing results in the literature for supersymmetric anharmonic oscillator, we investigated SUSY breaking in a model with degree-five superpotential and a shape invariant (Scarf I) superpotential. We then moved on to simulating the interesting case of models exhibiting certain type of $PT$-invariance. We simulated these models for various values of the parameter $\delta$ appearing in the $PT$-symmetric theory, without violating the reliability criteria for Markov chain Monte Carlo. Our simulations indicate that non-perturbative SUSY breaking is absent in quantum mechanics models exhibiting $PT$ symmetry. 

For the case of models with $PT$ symmetric superpotentials, an investigation that takes care of arbitrary $\delta$ values needs a simulation algorithm that can handle complex actions, such as the complex Langevin method. In Ref. \cite{Joseph:2019sof} it was shown that $PT$ symmetry is preserved in zero-dimensional supersymmetric theories. Recently, this study was extended to supersymmetric quantum mechanics models with arbitrary $\delta$ parameter \cite{Joseph:2020gdh} and there, with the help of complex Langevin simulations, it was shown that SUSY is preserved in these models. Our results in this paper complement these investigations.

It would be interesting to reproduce the conclusion given in Ref. \cite{Bender:1997ps} using non-perturbative methods such as Monte Carlo method or complex Langevin dynamics. There, it was shown that SUSY remains unbroken, using perturbative calculations, in a two-dimensional supersymmetric field theory exhibiting $PT$ symmetry. It would also be interesting to perform simulations in four-dimensional supersymmetric models exhibiting $PT$ symmetry and thus comment on the nature of the spectrum of the theory and implications to Higgs physics.         

\section{Acknowledgements}

We thank discussions with Raghav Jha and Arpith Kumar. The work of AJ was supported in part by the Start-up Research Grant (No. SRG/2019/002035) from the Science and Engineering Research Board (SERB), Government of India, and in part by a Seed Grant from the Indian Institute of Science Education and Research (IISER) Mohali. NSD would like to thank the Council of Scientific and Industrial Research (CSIR), Government of India, for the financial support through a research fellowship (Award No. 09/947(0119)/2019-EMR-I).

\bibliography{paper_refs}
\bibliographystyle{utphys.bst}


\end{document}